\documentclass[10pt,journal,compsoc]{IEEEtran}
\ifCLASSOPTIONcompsoc
  \usepackage[nocompress]{cite}
\else
  \usepackage{cite}
\fi
\ifCLASSINFOpdf
   \usepackage[pdftex]{graphicx}
\else
\fi

\usepackage{amsfonts}
\usepackage{amsmath}

\newcommand{\R}{\mathbb{R}} 

\newcommand{\BigO}{\mathcal{O}} 

\newcommand{\Paragraph}[1]{\vspace{0.01in}\noindent\textbf{#1}}
\usepackage{enumitem}

\newcommand{\toprule}{\hline}
\newcommand{\midrule}{\hline}
\newcommand{\bottomrule}{\hline}

\begin{document}
%
\title{Visualization of topology optimization designs with representative subset selection}
%
%
%
%

\author{Daniel~J~Perry,
        Vahid~Keshavarzzadeh,
        Shireen~Y~Elhabian,
        Robert~M~Kirby,~\IEEEmembership{Member,~IEEE,}
        Michael~Gleicher,~\IEEEmembership{Member,~IEEE,}
        Ross~T~Whitaker,~\IEEEmembership{Fellow,~IEEE,}
\IEEEcompsocitemizethanks{
\IEEEcompsocthanksitem V. Keshavarzzadeh, S. Elhabian, R. Kirby, R. Whitaker are with the SCI Institute, University of Utah, Salt Lake City, UT, 84112.\protect\\
E-mail: \{vkeshava@sci,shireen@sci,kirby@cs,whtaker@cs\}.utah.edu\protect\\
\IEEEcompsocthanksitem D. Perry work done while at the SCI Institute, University of Utah.\protect\\
E-mail: dperry@cs.utah.edu\protect\\
\IEEEcompsocthanksitem M. Gleicher is with the Department of Computer Sciences, University of Wisconsin - Madison, Madison, WI 53706.\protect\\
E-mail: gleicher@cs.wisc.edu
}
\thanks{Manuscript received August 31, 2017.}}

\IEEEtitleabstractindextext{%
\begin{abstract}
An important new trend in additive manufacturing is the use of
optimization to automatically design industrial objects, such as
beams, rudders or wings.  {\em Topology optimization}, as it is often
called, computes the best configuration of material over a 3D space,
typically represented as a grid, in order to satisfy or optimize
physical parameters.  Designers using these automated systems often
seek to understand the interaction of physical constraints with the
final design and its implications for other physical characteristics.
Such understanding is challenging because the space of designs is
large and small changes in parameters can result in radically
different designs.  We propose to address these challenges using a
visualization approach for exploring the space of design solutions.
The core of our novel approach is to summarize the space (ensemble of
solutions) by automatically selecting a set of examples and to
represent the complete set of solutions as combinations of these
examples. The representative examples create a meaningful
parameterization of the design space that can be explored using
standard visualization techniques for high-dimensional spaces.  We
present evaluations of our subset selection technique and that the
overall approach addresses the needs of expert designers.

\end{abstract}

}

\maketitle

\IEEEdisplaynontitleabstractindextext

%
\IEEEpeerreviewmaketitle

\IEEEraisesectionheading{\section{Introduction}\label{sec:introduction}}
%

\begin{figure*}[tb]
\centering
\includegraphics[width=.8\textwidth]{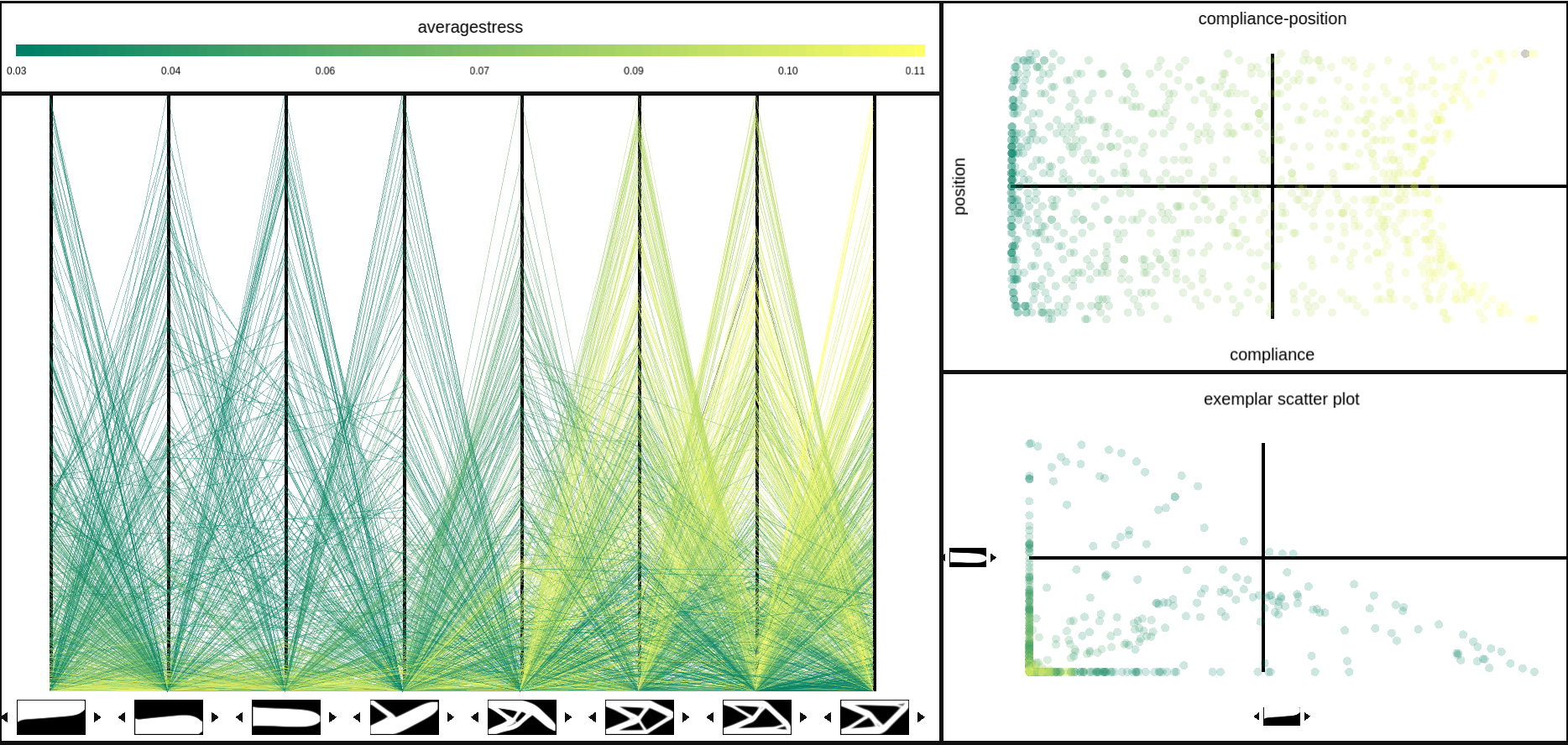}
\caption{%
A structure topology optimization (STO) design dataset shown using a representative subset based visualization (left side), along with supporting linked views (right side).  
The main (left) view shows a subset chosen for optimal representation with the remaining data points represented by the lines in the parallel coordinate view (colored by average stress of the design).
The top right shows the relationship between a single parameter, position, and the compliance of the design.
The bottom right view shows a close-up view between two of the example-axis, allowing a more detailed inspection of any relationships.
In our expert case study, this view allowed the domain experts to identify an interesting relationship between position, compliance, and the corresponding shapes by viewing all this information together.
}
\label{fig:teaser}
\end{figure*}

\IEEEPARstart{D}{esigners} in various industries (e.g.,  machinery construction, 3D
printing, automotive and aerospace) are often tasked with creating
products that meet certain constraints, such as making the most efficient use of
material (and reducing weight or manufacturing costs), achieving a specified level of thermal conductivity, or bearing a specified mechanical load. 
Meanwhile, additive manufacturing (AM) has removed the limitations of
traditional machine-tool-based manufacturing and opened the doors for
radically new physical designs.   
In AM, material is typically distributed
within a particular design space, for a given set of design parameters
and constraints, to optimize product performance criteria. 
As an emerging component of the engineering
design process, \textit{structural topology optimization} (STO) has
offered a principled approach for addressing sets of competing design goals 
by optimizing material distribution in the given design
space subject to problem-specific parameters and constraints
\cite{bendsoe1988generating,bendsoe2003topology}. Recent advances in manufacturing
technologies have made the STO approach to product design even more
promising and important.  For example, additive manufacturing can more directly map
such optimal material layouts to physical realizations
\cite{wu2016system,schramm2006recent}. 
%
%

While material placement is found automatically, other
design parameters, such as boundary conditions, external force
parameters, etc., cannot be set arbitrarily because the number and
range of values of these parameters can significantly impact solution
accuracy and convergence. Designers using topology optimization must also
understand the implications of different parameters and other physical
constrains to properly select designs and incorporate potentially
innovative ideas from the optimization solutions into more traditional
design pipelines. 
We refer to this choice of parameters and resulting STO design possibilities as the \emph{design space}.
However, \textbf{building a more holistic understanding is
challenging} due to the \textit{large design space} resulting from large number
of possible parameters and the \textit{high dimensionality} of the space in
which optimized topologies live (e.g., the space of all possible designs). 
Furthermore, for domain experts there
is a \textit{surprising variation} in designs from small changes of the design
parameters that are often \textit{difficult for a designer to anticipate}, thereby
complicating the process of understanding optimal designs and
incorporating human experience and knowledge into the engineering design
pipeline.  

The nonlinearity and nonconvexity of the topology optimization problem 
limit the availability of closed-form, analytical relationships between input parameters
and optimal designs \cite{ulu2016data}. Therefore,
ensembles of optimized topologies are often generated to aid in
exploring the underlying high-dimensional solution space. Despite the
computational burden of obtaining each structurally optimal topology,
typically a sufficiently space-filling sampling of the design
parameters space is used to find an ensemble of representative design
solutions. Exploration of such an ensemble can be performed by viewing
ensemble elements, i.e., optimized topologies, each in turn. However,
this task can become \textit{labor-intensive}, \textit{time-consuming}, and \textit{make
gleaning general ensemble trends difficult}. 

One typical approach to
analyzing ensembles in this setting, in hopes of understanding the
full space, is to deploy a dimensionality reduction technique, e.g.,
principal component analysis (PCA) \cite{ulu2016data}, to reduce the
associated degrees of freedom (i.e., topology space dimensions) in
representing each ensemble element. This type of analysis is
convenient and effective in terms of representation error (being optimal
in the $\ell_2$ reconstruction error), but the resulting basis functions are 
often difficult to understand in terms of the original
designs.  Furthermore, the reduced dimensions often do not provide an
adequate representation of important {\em features} in the data
(e.g., they are usually abstractions summarizing major trends rather
than enumerating all or most possibilities).  However, feature
coverage information, which relates information about the individual
ensemble elements, is important because they contain information about
design details and strategies (often tied to the design parameters and
application-specific performance criteria) that are often critical to
an engineers understanding of the design space.

To summarize, some of the \textbf{problems} encountered by a designer using STO which we plan to address include,
\begin{itemize}[noitemsep]
	\item high dimensional design space due to a large number of parameters and variables,
	\item unintuitive connection between design parameters and resulting optimal (non-unique) designs, and
	\item lack of interpretability in traditional dimension reduction, such as PCA.
\end{itemize}

In this paper, we propose a new visualization technique that is meant
to address the challenges of the topology optimization visualization
problem.  The proposed method uses an
automatically chosen subset of the ensemble itself as both a
\emph{summary} of the dataset and a \emph{basis} from which to view
the remaining data points.  This summary provides a better
representative coverage of the features in the dataset than
alternatives such as PCA.  Using the subset as a \emph{basis}, the full
dataset is represented as a combination of these example points,
resulting in a new coordinate system that can be visualized using
conventional visualization techniques.  
To make this clear, an example of our final visualization tool is shown in Figure~\ref{fig:teaser}, although we will explain the various components of the tool throughout the paper.
The choice of the subset and the method of
combination become important in this context, and we provide some
methodology for those choices. This results in a novel view of the
topology optimization datasets that has the summarizing
characteristics favored in PCA but with a basis set that is easy to
relate back to the original data ensemble and design space.
We evaluate the proposed approach by
considering several case studies as part of a visualization system, as
well as specific studies validating the subset decisions. Other
visualization work has examined the individual design visualization
\cite{wu2016system}, but to our knowledge, no one has previously
explored visualization paradigms to understand the design space of an
ensemble of topology optimization solutions. While our focus is
topology optimization design spaces, we note that the proposed
approach has potential implications in other domains with similar
properties, such as simulation ensembles, large image datasets, and
large text corpora.
This work makes the following novel \textbf{contributions}:
\begin{itemize}[noitemsep]
	\item characterizing and addressing the problem of
          visualization of topology optimization design, 
	\item proposing a novel, subset-based method for visualizing
          high-dimensional data to aide in topology optimization,  and
	\item novel evaluation methods and evaluation results in
          comparing subset selection for optimal design visualization.
\end{itemize}

\section{Topology optimization for industrial design}

Structural topology optimization (STO) has emerged as a powerful tool in designing various high
performance structures, from medical implants and prosthetics to jet
engine components
\cite{sutradhar2010topological,reist2010topology,zhu2016topology}. 
STO is different from shape optimization in which the topology is
predetermined and only the boundaries are optimized
\cite{chen2017design}. 
Therefore STO can be used to generate totally
new design strategies with significantly better performance. 



STO has received significant attention in the past couple of decades
due to the popularization of additive manufacturing \cite{brackett2011topology}. 
Some complicated designs may not be
realizable with traditional manufacturing techniques such as casting
or machining, which have manufacturing constraints such as tool access
or shape limitations associated with a molding process. In those
cases, there has to be a compromise between optimal design and
manufacturable designs. AM tackles these practical limitations and is
capable of producing designs with complicated geometries and/or topologies. In addition,
AM can be much faster than traditional techniques. Therefore, the
combination of STO as the computational engine and AM as the
manufacturing procedure is perceived as an emerging method for
producing high performance designs that have not been realizable
heretofore~\cite{lazarov2016length}.

For the purposes of simplifying this discussion we consider a certain 
class of prototypical STO problems in which 
the optimization finds the best material layout in the design
space in order to maximize the stiffness (or equivalently minimizing the
deflection or compliance) of the structure subject to a material volume
constraint. The response of the
structure for a set of loading and boundary conditions is typical computed/simulated via
finite element method (FEM). The topology optimization yields a
solution in the form of binary maps that indicate material placement.

\begin{figure}[!h]
\centering
\includegraphics[width=.6\columnwidth]{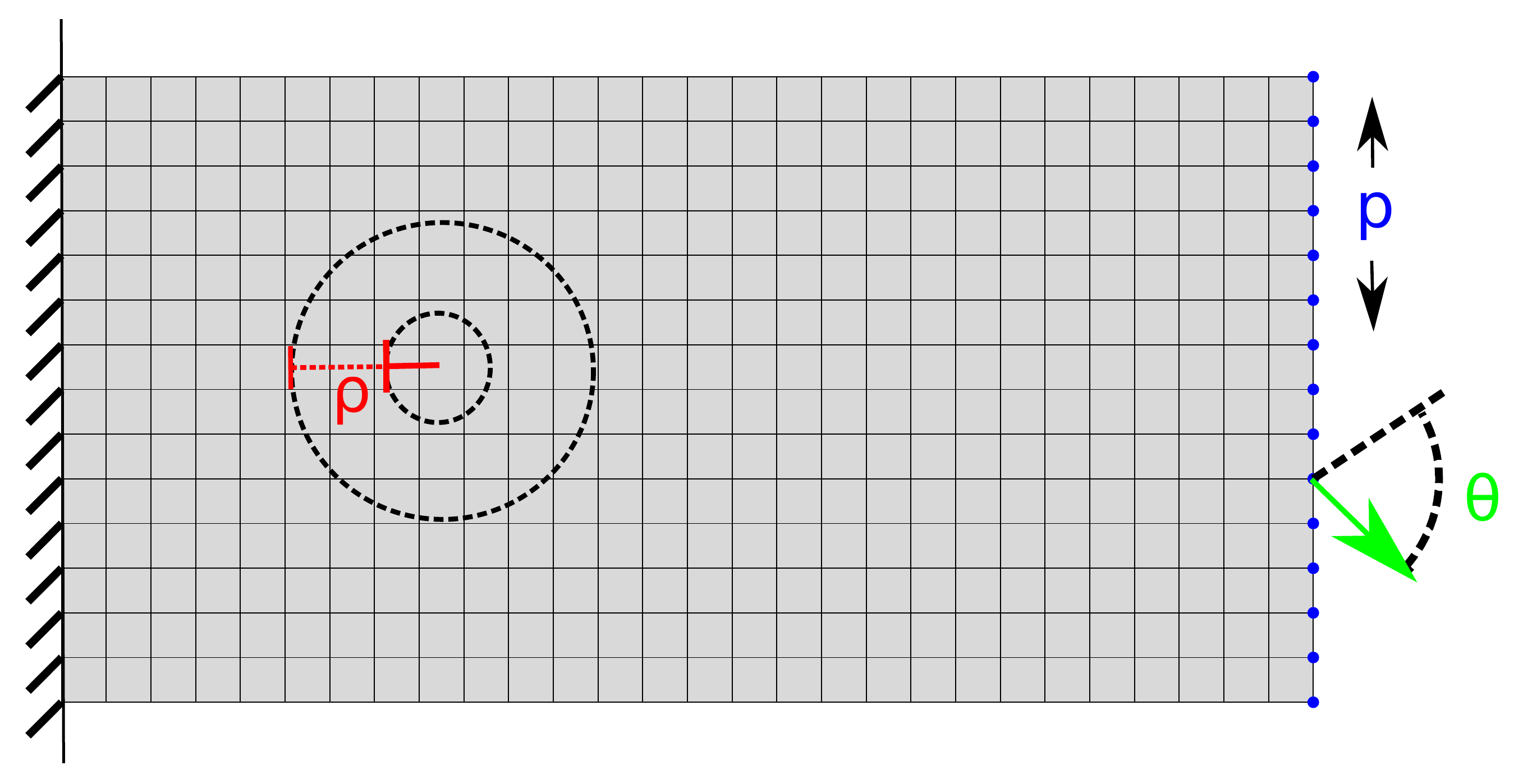}
\includegraphics[width=.38\columnwidth]{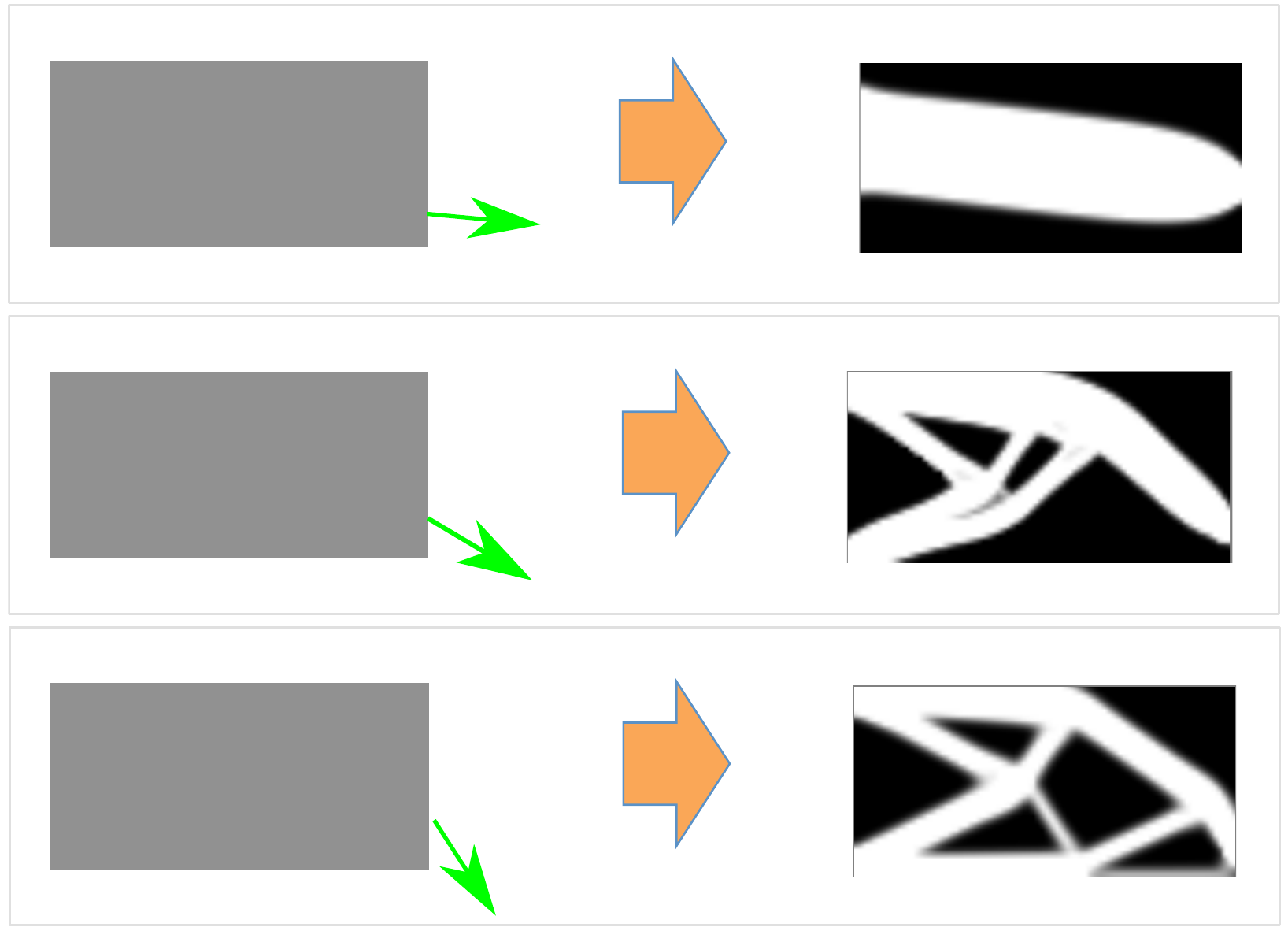}
\caption{
\textbf{STO design space.} 
\textbf{Left:} The loading on the right edge is parameterized by a vertical position, $p$, and a loading angle, $\theta$, and the filter size, $\rho$, which controls the length scale in the topology optimization.
\textbf{Right:} Some of the more obvious connections between the force angle parameter choice, $\theta$  and resulting STO  designs.
A shallow angle of force encourages the material to be centrally located into what we refer to as a \emph{beam} or \emph{beam-like} structure (right-top).
As the angle changes to become more inclined, the resulting design to includes smaller and smaller connections, ultimately becoming \emph{lattice-like} because the collection of smaller beams form a kind of lattice (right-middle,-bottom).
}\label{fig:struct}
\end{figure}

Our experiments and evaluations will use solutions from the design of a cantilever beam that is
clamped on the left edge and is under a static load on the right edge.
The variable loading is parameterized by the vertical loading
position, $p$, and the angle ($\theta$) of loading direction. The STO
is further constrained by a filter size parameter, $\rho$, which
determines the granularity of the material placements and effectively
controls the size of the structures that are allowed to form in the solution.
This design space is summarized on the left side of Figure~\ref{fig:struct}.

Some specific design results are shown in the right side of Figure~\ref{fig:struct}.
In order to facilitate discussion, we will introduce some colloquial terminology we will use throughout the paper to describe various aspects of designs in this space.
We will refer to designs which include a primary structure as a \emph{beam} or \emph{beam-like}, similar to the example on the top of Figure~\ref{fig:struct}, and we will refer to examples with a collection of smaller connecting structures as a \emph{lattice} or \emph{lattice-like}, with one example shown on the bottom of Figure~\ref{fig:struct}.
Along those lines the design in the middle of Figure~\ref{fig:struct} we will describe as having both lattice-like features with a beam structure along the top side of the design.
Additional examples can be found in a later part of the paper in Figure~\ref{fig:subsets}, as well as in many of the visualizations throughout.


\section{Related work}
\subsection{Dimension reduction}
The proposed method falls in the general class of dimensionality
reduction (DR) techniques, but is designed to address specific
challenges of feature summarization and coverage in the topology
optimization context.
However, because it shares many characteristics with more general
DR we review that related work first to identify
specific differences and similarities to existing methods.  

An important aspect of our approach is to use examples from the dataset to summarize and represent the rest of the data.
The idea of using a subset for DR is important and has been explored in other forms \cite{marks1997design,paulovich2008least,chen2009exemplar,paulovich2010two,joia2011local,kim2016interaxis}.
In \cite{kim2016interaxis} they use examples to define the extents of an embedding space because the examples provide an undiluted representation of the underlying set.
In another set of examples, \cite{joia2015uncovering} and \cite{marks1997design} both use a subset of the dataset to represent the contents of data clusters or to identify points in a 2D embedded layout, because the subset presents a natural summary of data without any confusing abstraction or obfuscation.
The works \cite{paulovich2008least,chen2009exemplar,paulovich2010two,joia2011local} all use subsets because of their numerical closeness to the remaining data elements, allowing for the creation of powerful embedding methods.
Subsets are a powerful representations because they contain details of the dataset, they avoid abstraction, and when selected appropriately can accurately represent neighbors.

One important goal of this work, which differentiates it from most DR
approaches, is that we require a basis or axis after the DR
transformation that easily relates back to the original data and
provides explicit information on the existence of \emph{human-identifiable features} (see Section~\ref{sec:subsets} for a more precise treatment of this requirement).
For the specific application to STO designs presented here, the axis should be easily related back to
examples of STO designs, because these are physically meaningful and
optimal, and become important for connecting designs and parameters.
From the above discussion subsets are clearly an excellent choice to satisfy this interpretability constraint of STO summarization.

One of the most widely used DR techniques is principal component analysis (PCA) \cite{jolliffe2002principal}.
PCA seeks a linear projection to a lower dimension that maximally preserves variance of the dataset.
While PCA is optimal in linear representation, the resulting basis is
often quite difficult to interpret and frequently has only a vague
relationship to the individual input data samples.
In this paper we will examine PCA with respect to the STO application
domain and show examples from an expert case study where using the PCA
as the basis made otherwise simple tasks quite difficult. 
These same problems can also be found in many {\em optimal basis} or
embedding techniques, such
as independent component analysis (ICA)
\cite{hyvarinen2004independent} and multidimensional scaling (MDS)
\cite{kruskal1978multidimensional, ingram2009glimmer}. 

There are many nonlinear DR techniques, including nonlinear extensions
of PCA, such as kernel PCA \cite{scholkopf2002learning}, or nonlinear
geometric DR methods, such as Isomap \cite{tenenbaum2000global} and
local linear embedding (LLE) \cite{roweis2000nonlinear}, that have
been used to better model the geometric relationship of the data.
While these approaches have arguably better representations of data on
nonlinear manifolds, they do not directly address the interpretation
problem, and, indeed, the resulting parameterizations often have no
physical interpretation.
Likewise, probabilistic nonlinear DR methods, such as stochastic neighbor
embedding (SNE) \cite{hinton2002stochastic}, its t-distribution
variant t-SNE \cite{van2008visualizing}, and parametric embedding (PE)
\cite{iwata2007parametric}, provide no clear interpretation relative
to the original ensemble.  

Improving DR from an interpretation standpoint has also been addressed in
the literature \cite{coimbra2016explaining,martins2014visual,joia2015uncovering}.
While these methods aid in the use of standard DR techniques, they do not
address the need for the axis to directly encode related human-identifiable features.
For example, \cite{coimbra2016explaining} and \cite{martins2014visual}
both provide important tools for either finding the right point of
view of the DR scatterplot or enhancing the choice of DR parameters
for existing systems.  Unfortunately, both cases, as with PCA, do not address the underlying problem that the DR axis obfuscate many of the human-identifiable features.
\cite{joia2015uncovering} more closely addresses our needs by automatically selecting a subset to represent clusters in the DR.
We compare to this general idea of clusters being representative of the data by using a k-medoids clustering for subset selection in some of our tests.
The proposed approach is different in that it uses the subset, however
it is determined, as the axis of the DR, while they perform a DR and then use a subset to represent resulting clusters.
Additionally, the subset selection method from
\cite{joia2015uncovering} has some fundamental problems which we
directly address in our subset methodology.    Their method first computes an singular value decomposition (SVD), closely related to PCA, and then finds points similar to the singular vectors (PCA modes).
While this will work for specific datasets where points \emph{happen}
to lie near modes, \emph{this will not always be the case} (this is
similar to expecting the points to always lie near an average of the
points). We instead propose selecting a subset that directly minimizes this representation error.

\subsection{Subset-based visualization}
The proposed work is strongly motivated by previous visualization
research that has demonstrated the effectiveness of data subsets to
understand the structure of a population.  For instance, 
\cite{chen2009exemplar} uses greedy residual-based sampling to
select the subset and PE, referred to above, to create two-dimensional
embeddings.  

Another set of related techniques is least square projection (LSP)
\cite{paulovich2008least} and part-linear multidimensional projection
(PLMP) \cite{paulovich2010two}, that use a subset of the data to learn
a linear low dimensional embedding.  This is then applied to the full
dataset.  LSP is motivated by increasing the DR quality by limiting
the learned projection to a set of control points, while PLMP is
primarily motivated by computational challenges for extremely high
dimensions datasets.  LSP and PLMP both use clustering methods
(k-medoids and bisecting k-means) to select the controls points.  To
understand the effectiveness of this, 
we evaluate k-medoids as a method for selecting a visualization subset and
compare it to several other representation-motivated methods in this paper.  Both
LSP and PLMP result in a lower dimensional space whose axis are
weighted combinations of the subset, similar to PCA.

Local affine multidimensional projection (LAMP) \cite{joia2011local}
provides for user selection of the subset and then uses that subset to
produce a locally affine projection for the full dataset.  Another
interactive technique was presented in \cite{kim2016interaxis}, where
the user can interactively select basis extents from specific samples
in the dataset, for the purposes of controlling how the scatter plot
data is displayed.  The proposed method differs from both in that the
exemplars themselves are the axis and are selected automatically
(rather than by the user), allowing it to scale to large subsets and
large ensembles.
Additionally, the proposed method automatically selects the subset
based on some specific motivating criteria.  Another point of difference is
that \cite{kim2016interaxis} is focused entirely on a 2D scatter plot
embedding, while we have used the proposed method almost entirely in
higher-dimensional representations using alternative visualization
strategies, such as parallel coordinates.

Design galleries \cite{marks1997design} presents a 2D embedded layout and summarizes visual content of clusters or positions in the embedding using small multiples of a subset of the full dataset.
This approach to using subsets for summarizing feature content of data
points is quite effective, but is limited to 2D and 3D embeddings,
which are often insufficient to adequately represent the high-dimensional data.
This same limitation of embeddings also applies to many of the DR approaches described above.
In the proposed approach, the dimensionality is only limited by the
number of examples that can be shown in a row at the chosen screen and
image resolutions.

\subsection{Structural topology optimization analysis}
Because we are not contributing directly to the methods of solving STO
problems, we restrict our discussion to relevant work that analyzes or
visualizes this type of data (for a more complete review of STO in
general see, for example, \cite{bendsoe2003topology}).  In
the domain of visualization, \cite{wu2016system} introduced a complete
system for finding and visualizing individual STO solutions.  They
present an impressive array of STO design spaces and visualization
results, however their work focuses on the visualization of a single,
complicated design space.  Here, we are focused on understanding an
ensemble of designs found by sampling different points in the
parameter space or by different results from a stochastic process.

The work in \cite{ulu2016data} also has some overlap with this paper, 
because they use ensembles of designs to learn data driven models,
relating the parameters to design solutions via a neural network.
Their underlying motivation is to reduce the computational burden of
finding new design solutions in the large design space.  However,
rather than learning a data driven relationship, we are instead
interested in visualizing the data in such a way to allow the human
designer to understand that relationship, and then apply that
knowledge in future design problems.
Additionally, \cite{ulu2016data} relies on PCA as the primary DR stage of their pipeline.
This makes sense for a purely data-driven approach which does not
address interpretability, and therefore has the previously described limitations.

\section{Subset-based visualization of topology optimization designs} 
\label{sec:proposed} We propose a technique for
visualizing a large collection of high-dimensional designs by first selecting a representative subset,
using that subset as the visual basis, encoding the remaining data as
linear combinations of those elements, and then displaying the
resulting subset and encodings in standard visualization tools such as
parallel coordinates \cite{inselberg1990parallel} (or star coordinates
\cite{kandogan2000star,kandogan2001visualizing}, scatter plot matrices
\cite{liu2015visualizing}, etc.).

\begin{figure}[tb]
 \centering 
 \includegraphics[width=0.8\columnwidth]{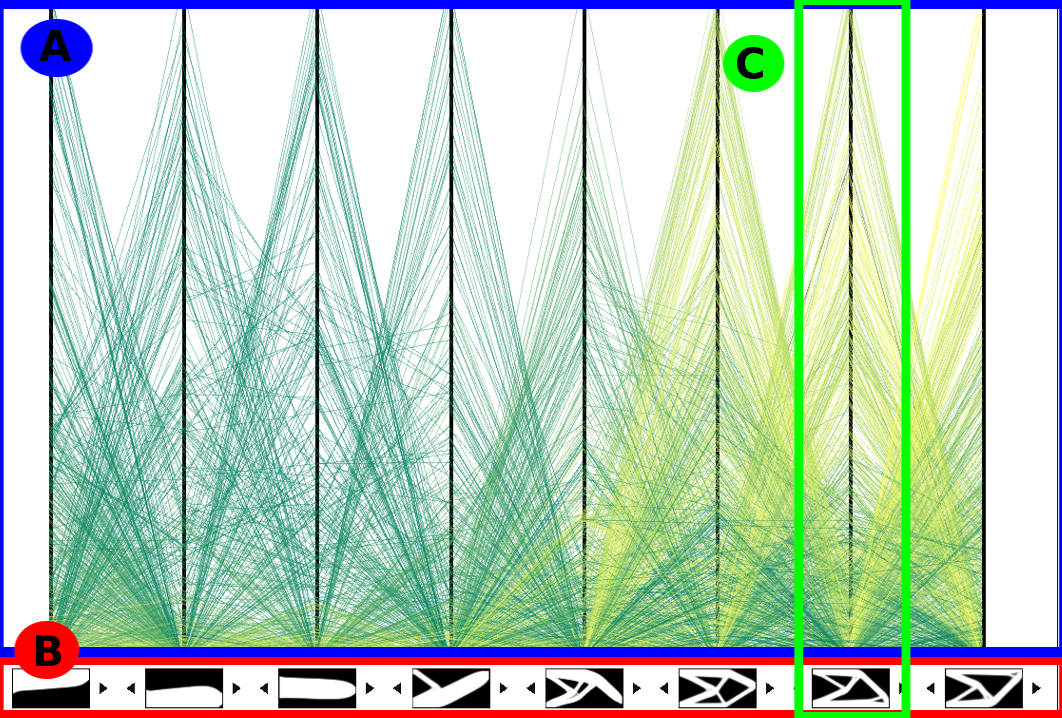}
 \caption{ \textbf{Diagram of the components of the proposed
     subset-based visualization using parallel coordinates.}
   Component \textbf{(A)} shows the main parallel coordinate display
   that plots the computed weights of all data points on the chosen
   subset, by selecting and filtering this view, one gains an idea of
   what portion of the population is similar to the coordinate bases.
   Component \textbf{(B)} shows the subset axis figures. This subset
   represents the visual features of the dataset, and viewing the
   weights of the remaining points provides some intuitive idea of
   what each of the individual shapes consists of by considering the
   larger weights.  Component \textbf{(C)} highlights a single
   coordinate axis, where the line positions along this axis are the
   weights of all points related to the highlighted subsample element.
 }
 \label{fig:components}
\end{figure}

An example is shown in Figure~\ref{fig:components}, where we have
highlighted the main components of the subset-based visualization.
The two main components are the subset display at the bottom (marked
\textbf{(B)}), and the parallel coordinate plot of the weights
relating the other data points to that subset (marked \textbf{(A)}).
Because these are the main components of the visualization, careful
selection of the type of subset and the properties of the weight
relationship becomes very important.
We additionally highlight a single coordinate axis in \textbf{(C)}, with line positions along the axis correspond to reconstruction weights of all points using that element.

\subsection{Representative subset} \label{sec:subsets}
The subset selection method can be motivated by different goals.  For
example, if we believe that data naturally stratifies into groups, a cluster-center-based
subsample might be appropriate.  In contrast, a random
set of samples represents an unbiased subset.  Because the
typical screen size limits the number of basis elements that can be displayed
(see Figure~\ref{fig:components}) it becomes critical to choose a
subset that is both relatively small and that accurately represents
the entire set of input data.  For this reason, we formulate the
representation goals of the subset decision precisely and then
pursue optimum satisfaction of these goals.  We define a subset as
\textbf{representative} if \emph{linear combinations of that subset
  minimize the reconstruction error of the rest of the dataset}.

If we express the input data in matrix form, $\mathbf{X} \in \R^{d \times
  n}$, where each column $\mathbf{x}_i \in \R^d$ is a data point and
each row is a feature or dimension,
the subset-based  representation can be formulated so that $\mathbf{X} \approx \mathbf{R} = \mathbf{D}\tilde{\mathbf{B}}$,
where $\mathbf{R} \in \R^{d \times n}$ is the representation (or
reconstruction),  $\mathbf{D} \in \R^{d \times m}$ is a subset of the
data, and $\tilde{\mathbf{B}} \in \R^{m \times n}$ is the coefficients
(or weights) used to combine the subsets into the approximate   reconstructions.
To find the \emph{most representative} subset would mean simply
minimizing the difference between the representation $\mathbf{R}$ and
the input data $\mathbf{X}$, or  
\begin{equation}
\min_{\tilde{\mathbf{B}},\mathbf{D} \subset \mathbf{A}} E(\tilde{\mathbf{B}},\mathbf{D}),
\end{equation}
where, 
\begin{eqnarray}
E(\tilde{\mathbf{B}},\mathbf{D}) &=& \sum_i \| \mathbf{x}_i - \mathbf{D}\tilde{\mathbf{b}}_i \|_2^2 \\
&=& \|\mathbf{X} - \mathbf{D}\tilde{\mathbf{B}}\|_F^2,
\end{eqnarray}
in which we have used $\mathbf{D} \subset \mathbf{A}$ to denote that
$\mathbf{D}$ is a column subset of $\mathbf{A}$. 

\begin{figure}[tb]
 \centering 
 \includegraphics[width=\columnwidth]{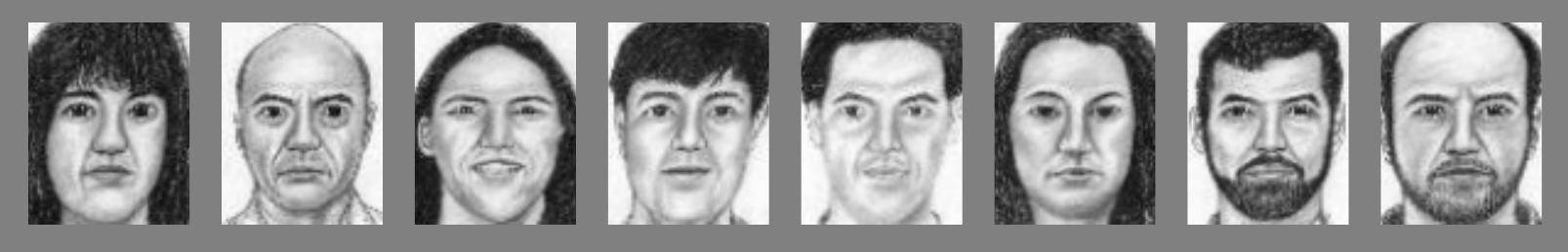}
 \includegraphics[width=\columnwidth]{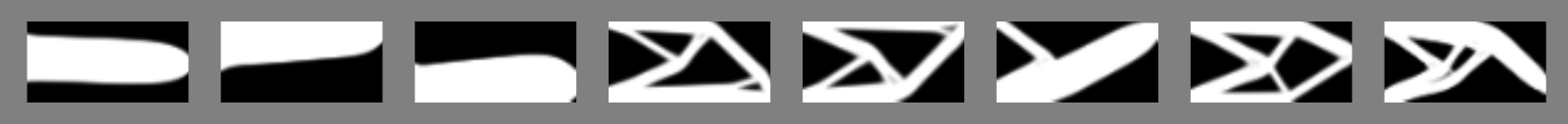}
 \includegraphics[width=\columnwidth]{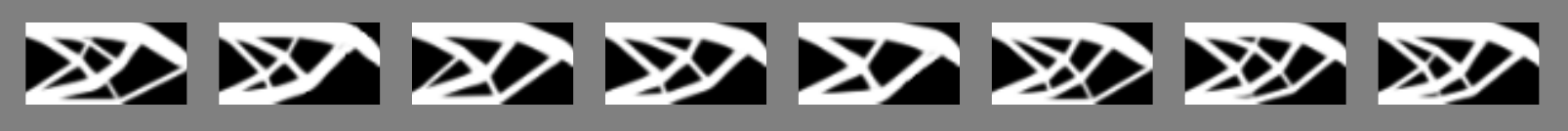}
 \caption{ \textbf{Representative subsets from face-sketch, STO D1, and STO D2.}
 \emph{Top:} \emph{representative} subset from a face-sketch dataset \cite{wang2009face},
 \emph{Middle:} from the STO D1 design dataset with varying parameters and constant initialization,
 \emph{Bottom:} from the STO D2 design dataset with constant parameters and varying initializations.
 }
 \label{fig:subsets}
\end{figure}

In considering our specific STO problem, we found that domain experts
are reliant on certain visual elements of the resulting designs which
we refer to as \emph{human-identifiable features}, or just
\emph{features}.   For STO designs these include, for example, the
number of internal holes, the number of thick beam structures, thin
beam structures, how latticed the design is, etc. 
Consider some of the designs shown in Figure~\ref{fig:subsets} (bottom two rows).
These features are particularly important when making \emph{connections between parameters and designs}.
This idea of \emph{human-identifiable features} is more general than
for just STO designs and can change, appropriate to the domain. 

As an intuitive example, we might consider the set of human faces that have
\emph{features} such as hair length, presence of facial hair, a hat,
earings, etc.,  as seen in the sketched human faces (from \cite{wang2009face}) in Figure~\ref{fig:subsets} (top).
These facial features are groups of data values (pixels) that
have a correlated behavior across multiple images.   For instance,
longer hair is seen as sets of values/pixels on either side of the
head that have dark values.   
Being able to identify the subset with the most representative
features reduces computationally to finding the subset whose features
(groups of pixels) maximally correlates with the rest of the dataset,
which matches well with our formulation above. 
In the estimation and machine learning literature, research have
observed that non-negative matrix factorization (NMF) has the property 
of extracting coherent features \cite{lee1999learning}, by insisting
that feature-capturing basis elements approximate the data by combining in a
{\em constructive} manner. 
In other words, the non-negative constraint on weights tends to
discourage the representation of data features through a
negative-positive combination of large numbers of basis elements (as
with PCA) and requires features (correlated values) to be represented
explicitly in the basis elements.   
This effect can be seen in well-known negative-positive weighted decompositions such as \emph{Eigenfaces} \cite{turk1991face}, which have face-like shapes but are difficult to interpret as clear facial features.
Additionally, we will describe in our expert case study in Section~\ref{sec:experts}, experiences where non-negative weights were easier for the participant to process mentally in terms of interpreting the combination of examples from the subset, reinforcing the importance of a non-negative weighting for relevant tasks.

With these details in mind, we further refine
our goal in finding a \emph{representative subset} which captures the
most \emph{representative features} by constraining the reconstruction
weights, $\tilde{\mathbf{B}}$, to be non-negative. 
Thus, the non-negative formulation of the subset problem is:
\begin{equation}
\min_{\tilde{\mathbf{B}} \ge 0,\mathbf{D} \subset \mathbf{A}} E(\tilde{\mathbf{B}},\mathbf{D}).
\end{equation}

\subsection{Simultaneous subset and coefficient computation} \label{sec:gomp}
We now propose an approach for solving the optimization problem
presented in the previous section. We can rewrite the objection function as, 
\begin{eqnarray}
E(\tilde{\mathbf{B}},\mathbf{D}) &=& \|\mathbf{X} - \mathbf{D}\tilde{\mathbf{B}}\|_F^2 \\
&=& \|\mathbf{X} - \mathbf{X}\mathbf{B}\|_F^2 = \hat{E}(\mathbf{B},m),
\end{eqnarray}
where $\mathbf{B} \in \R^{n \times n}, \mathbf{b}_{i,:} =
\tilde{\mathbf{b}}_{j,:}$, and $\tilde{\mathbf{b}}_{j,:}$ is the
$j$-th row corresponding to the $j$-th data element in $\mathbf{D}$, 
the same as the $i$-th data element in $\mathbf{X}$. 
The smaller matrix $\tilde{\mathbf{B}}$ can be thought of as a
submatrix of the larger more sparse $\mathbf{B}$, where $m$
specific (nonzero) rows are preserved. 

Instead of building $\mathbf{B}$ from $\tilde{\mathbf{B}}$, we are
interested in solving for $\mathbf{B}$ directly (and therefore
obtain the embedded $\tilde{\mathbf{B}}$).  The absolute optimum of
the objective $\|\mathbf{X} - \mathbf{X} \mathbf{B}\|_F^2$ would result in
$\mathbf{B} = \mathbf{I}$. However, if we constrain or regularize
$\mathbf{B}$ so that it is row-sparse and non-negative, as described above, we can
recover an optimal $\mathbf{B}$ with the desired properties.  
The regularization approach leads to this formulation for subset selection,
\begin{equation} \label{eqn:obj}
\mathbf{B}^* = \arg \min_{\mathbf{B} \ge 0} \|\mathbf{X - XB}\|_F^2 + \lambda \|\mathbf{B}\|_{2,1}
\end{equation}
where $\|\mathbf{B}\|_{2,1} = \sum_i \|\mathbf{B}_{:,i}\|_2$ is a mixed $\ell_{2,1}$
norm that induces row sparsity, where only a subset of the rows has nonzero entries, and
$\mathbf{B} \ge 0$ means element-wise non-negativity ($b_{i,j} \ge 0, \forall i,j$).



The resulting $\mathbf{B}$ matrix encodes both the subset selection (corresponding to nonzero rows), and the weights connecting the rest of the dataset to the chosen subset (the values of the matrix).
This is a convenient result as both the subset selection and weight computation occur simultaneously.
This problem is known as a \emph{grouped lasso} problem, and many
solvers exist for efficiently estimating the matrix $\mathbf{B}$
\cite{yuan2006model,boyd2011alternating,deng2011group}. 
We have found the method choice does not affect the resulting subset
significantly, and have settled on using the fast grouped orthogonal
matching pursuit (GOMP) solver\cite{kim2012group}, which was used in
all results presented in this work. 

\subsection{Subset size}
One important parameter to consider is the size of the subset.  This
parameter presents an important tradeoff.  A larger subset will
provide the opportunity to represent more unique \emph{features} in
the dataset, but a large enough subset could become cluttered, redundant, or even
confusing.  However, we have found that the screen space required for
the proposed layout in Figure~\ref{fig:components} is ultimately the
limiting factor in the subset size, and so our recommendation is
to use as many as the screen layout will allow, within reason.  Another
factor, which we have found useful in the decision, is to measure
reconstruction error. For some datasets viewing the reconstruction
error as a function of subset size can be helpful in determining a
good tradeoff.  In all of the experiments here, we use a subset of
size 8, because we found this was appropriate for the screen displays
used in our experiments and found it sufficiently large to capture the
most important variations in the dataset. 

\subsection{Visual encoding}
The method above places each design as a coordinate point in a
{\em medium-dimensional} (e.g., $d=8$) space that is optimized to be representative
and capture significant features.  We have investigated a
variety of methods for visually encoding the resulting
\emph{representative subset coordinates}, including scatter plot
matrices, parallel coordinates, and star coordinates.  Because of the
non-negative combinations, the star-coordinate plot was promising.
However, the documented ambiguities in the star-coordinate
visualizations prevented sound exploration of the data and pattern
discovery \cite{liu2015visualizing,lehmann2013orthographic}.  The
scatter plot matrix was difficult to use because of the large number
of zero coefficients in the non-negative case, resulting in individual
scatter plots that had points mainly clustered near the origin---the
properties of which are unclear, but an area of ongoing consideration/investigation.
Consequently, for this work we found the parallel coordinate
visualization of the subset coordinates to be more effective, because
it makes more visually apparent the coefficient relationships, better
disambiguates data points, and, with the addition of interactive
selection and filtering, makes it easier to find important
relationships among data points.  These motivations combined result in
the proposed visualization shown in Figure~\ref{fig:components}.

\subsection{Linked views}
As described previously, the topology optimization designs are useful
for STO task only when visualized coincident with parameters
that gave rise to each design and corresponding scores related to
expected functionality (e.g., maximum stress).  Therefore, we augment
the subset-based parallel coordinates view with a scatter plot showing
combinations of the various parameters and scores.  This additional
view is then linked to the parallel coordinates view by using the same
coloring, point selection, and filtering effects in both views.

We include two types of linked scatter plots.  The first is a detailed
view of the two selected axes from the parallel plots that we call the
\emph{exemplar scatter plot}.  The second is a scatter plot that can
use several options chosen by the user, from a t-SNE or isomap
embedding of the dataset, to a scatter plot of specific parameters
(angle, position, filter size) or score values (max stress, average
stress, compliance).  This additional visualization is a traditional scatter
plot of parameter data, augmenting the subset display, and therefore 
the concerns about traditional DR techniques are not directly relevant.  
For example, some DR methods make cluster
identification simple, and connecting that view to the proposed
subset-based view would allow for fast interrogation of cluster
attributes by visualization of the subset representations.

\begin{figure}[tb]
 \centering 
 \includegraphics[width=0.48\columnwidth]{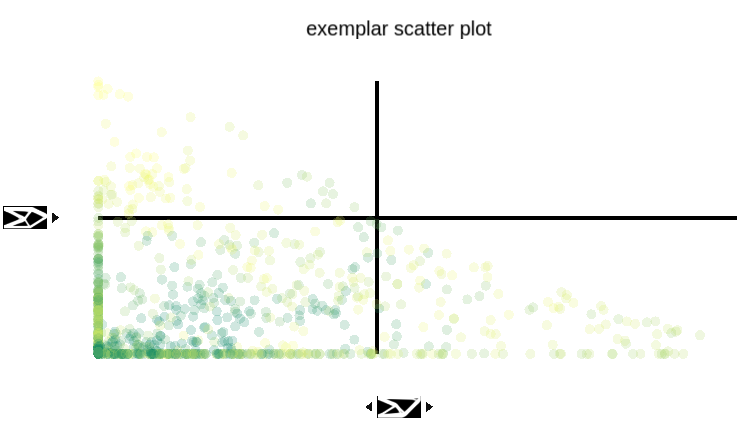} 
 \includegraphics[width=0.48\columnwidth]{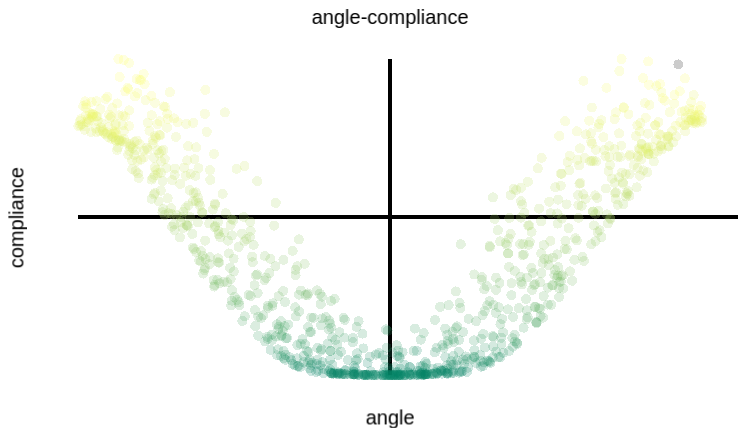} \\
\phantom{A} \hspace{10mm}	\textbf{Exemplar scatter plot} \hfill \textbf{Parameter scatter plot} \hspace{10mm} \phantom{A} \\
 \includegraphics[width=\columnwidth]{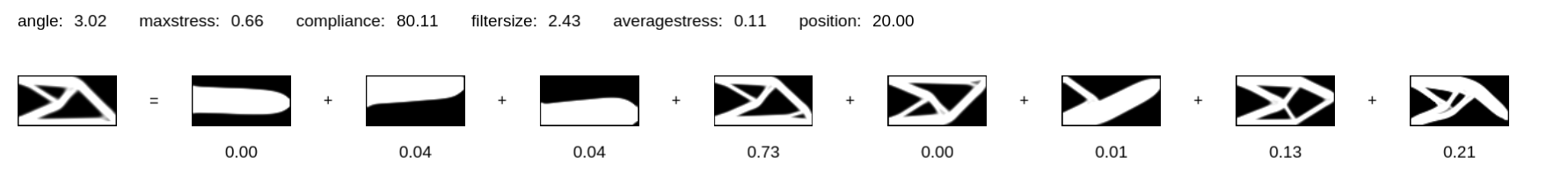} \\
	\textbf{Example of the detail view showing a specific design}
 \includegraphics[width=\columnwidth]{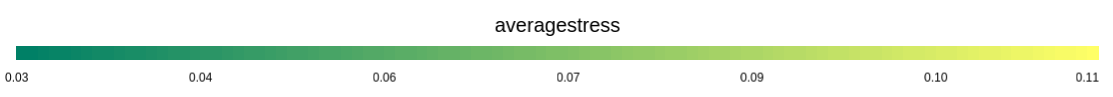} \\
	\textbf{The colormap used throughout the examples}
 \caption{
\textbf{Linked scatter plot and detail view.}
\emph{Top:} The exemplar scatter plot allowing detail view of two parallel coordinate views, and the second scatter plot view that can be set to a scatter plot of parameters (shown) or an embedding like t-SNE, Isomap, etc. as needed.
\emph{Middle:} An example of the detail view of a single data point including specific parameter values and weights used in a linear reconstruction.
\emph{Bottom:} The colormap used in all examples, representing low to high values of the user-selected parameter.  For this figure points are colored according to \emph{average stress} of the corresponding design.
}
 \label{fig:sample}
\end{figure}

Additionally, we include a linked detail view that allows the user to
view a specific data point and related details such as weight values. 
Figure~\ref{fig:sample} shows a few samples of each of these linked
views, and Figure~\ref{fig:teaser} shows a sample of them combined as
a single visualization tool. 

\section{Evaluation}
We assessed the method in two parts.  The first part, \textbf{subset
  selection evaluation}, focuses on evaluating the proposed subset
selection method, using a quantitative analysis of \emph{feature
  coverage} for each subset selection approach considered, and a
quantitative assessment of \emph{representation} error in a combined
subset and weight computation. The second part, \textbf{full system
  evaluation}, focuses on the overall visualization with a
quantitative analysis of \emph{successful task completion} for the
domain experts and a qualitative assessment of \emph{usage} from a set
of domain expert case studies.
In addition to the evaluation, we include a final clarifying example to show how the full systems works on even very challenging datasets.

\subsection{Subset selection evaluation} 
\label{sec:subset:eval} 
An important characteristic of the resulting subset is including
examples of a broad range of user observable features of the dataset
elements.  The importance of this \emph{feature coverage} arises from
the need to represent a variety of points outside the chosen subset
that will each include a \emph{potentially different mixture of those
  features}.  Likewise, any features not represented in the subset by
definition will not be available in the visualization.

We consider our approach to compare feature coverage as a \emph{novel contribution}, as we are unaware of previous work that has used feature coverage tests in this way to evaluate subset selection.
This approach to measuring the level of coverage of the observable features allows a more direct, objective view of how well the subset represents what a user would see had they looked at all the elements of the dataset one by one.

The novelty is most easily seen in image-based datasets where observable features are difficult to extract automatically from the data elements.
Specifically we select the subset using only the data elements, ignoring any manually-generated labels that would not be available in a completely new, unlabeled dataset that we would want to process.
After selecting the subset, we use human-labeled visual features to determine quantitatively how well the subset covers the observable features -- if the subset has a high coverage percentage, then it represents the dataset well, and if it has a low coverage percentage, then it does not represent the full dataset well.
This comparison approach provides a strong indication of the quality of the subset from the point of view of representing the observable features in the full dataset.

\subsubsection{Methods}
We compared the proposed subset selection method, solved using
non-negative group orthogonal matching pursuit (GOMP-NN), with the
deterministic form of interpretable decomposition subset method (ID)
\cite{cheng2005compression,golub1965numerical}, the k-medoids method
(KM) \cite{park2009simple}, and uniformly random subsets (RAND).  The
formulation of the proposed GOMP-NN approach was described previously
in Section~\ref{sec:gomp}.  ID is an appropriate alternative because
it also simultaneously computes the weights, but it instead uses a
pivoted-QR-type \cite{golub1965numerical} subset selection,
computing both positive and negative weights.
KM (and related k-means)
is a very well-known approach to clustering and/or quantizing a set of
data, which also identifies cluster centers.
Because KM does not explicitly prescribe a weight computation, we will
look at both non-negative (KM-NN) and positive-negative weight options
(KM-PN).  RAND is a common method to subsample a dataset in an
unbiased way.  RAND also does not prescribe a specific type of
weighting, so we compare against both options (RAND-NN and RAND-PN).
One critical difference between GOMP-NN and either KM-NN or RAND-NN is
that the non-negativity also drives the subset selection in GOMP-NN.
In contrast, for KM-NN or RAND-NN, the nonnegative weights can only be 
computed after the subset selection.

In terms of asymptotic runtime, for a dataset with $n$ elements of $d$
dimension, GOMP is $\BigO(mn^2d)$, ID is $\BigO(mnd)$, KM is
$\BigO(mndt)$, and RAND is $\BigO(1)$, where $t$ is the number of
iterations to convergence for KM.  For datasets where $n$ is the
largest dimension, GOMP is the most expensive, followed by KM, and
then ID.  RAND is the fastest with a constant cost.  However, the STO
design datasets have a data dimension much
larger than the number of points ($d > n$).  Because of the high cost of solving
for a single design, this situation will be common for general STO
design ensembles.  As the dimension of the design space increases, the
cost for each solution will also increase, limiting the number of
samples that can be solved.  This indicates that asymptotic cost in
$n$ will become less of an issue as larger systems are solved.  From
this vantage point, the runtime differences between GOMP, ID, and KM
become less pronounced.  KM is the only non-greedy method in terms of
subset selection.  For example, choosing $m$ verses $m+1$ subset
elements can result in two different subset choices, while for both
GOMP and ID the $m+1$ subset is the same as the $m$ subset except for
the additional subset element found in the latest round.  This is
important when evaluating different size subsets because as the subset
grows ID and GOMP both only require an additional sample to be found,
where for KM the entire subset must be recomputed.  Overall, for the
STO design datasets we have not found a significant difference in
runtime cost among methods (excepting RAND), and so limit the runtime
discussion to this summary.


\subsubsection{Datasets} \label{sec:subset:datasets} 
We include an
evaluation of subset method results on two datasets.  One,
\textbf{CelebA}, is a database of celebrity face images used in
\cite{liu2015faceattributes}.  This dataset is of interest for the
current problem because of the intuitive nature of the dataset type
(everyone understands common face features).   Also, this specific
dataset contains a set of $40$ human-derived facial features amenable
to the comparison, such as ``mustache'', ``eye glasses'', or ``pointy
nose'', etc.  Evaluating this dataset also supports the claim that the
proposed technique is more general than only STO design
datasets.  The original dataset consists of $2 \times 10^5$ color face
images of size $178 \times 218$.  Because this dataset is not the
primary focus of the paper, we subsampled it both spatially (down
sample by $4$) and randomly in the number of images (to $2 \times
10^4$).  As the STO datasets are much smaller than this down-sampled
version, we consider this compromise acceptable.

The other, \textbf{STO D1}, is an STO design
dataset, generated by collaborators/coauthors, with a design space of
$40 \times 80$ pixels and $10^3$ 
samples, where the structure of interest is attached to the left edge
of the domain, and a point force is applied at some position $p \in
[-20,20]$ and angle $\theta \in [0,\pi]$ on the right edge.  The
optimization is further constrained by a filter size $\rho \in
[1.1,2.5]$ that controls how thin/thick structures are allowed in the
solution (see Figure~\ref{fig:struct}).  A Monte Carlo sampling among
these three parameters was used ($10^3$ samples), with the same
constant initialization.  Because this dataset did not have a
predefined set of human-defined features like CelebA, we generated
and evaluated them within a small-scale user study.  A sample of the dataset is
shown in the middle portion of Figure~\ref{fig:subsets}. 


\subsubsection{User-based evaluation of STO
  designs} \label{sec:userstudy}
To generate the human-defined features, we recruited one STO expert
and eight different nonexpert graduate students.  We asked the STO
expert to specify a set of physically relevant features to describe the
set of designs in a
way that a nonexpert could understand, including things such as number of
internal holes, number of external holes, thickness of the components,
etc.    We then asked
three of the nonexpert participants to generate a set of visual features
from their own assessment of the dataset.  The expert and nonexperts
each generated the feature set while being shown five random subsets
of size 25 (125 total samples) from the dataset.  One
of the resulting nonexpert lists was not a binary feature set (as per
the instructions) that had to be thrown out.  After verifying both of
the other nonexpert lists were binary feature sets (they could be
evaluated in an approximate ``yes'' or ``no'' answer for any example), we randomly chose
one of the remaining two nonexpert study for the remainder of the
study, so that comparison could be made across participants.

We then asked five of the nonexperts to assess which of the subsets
contained at least one instance of each feature in the expert list,
and four do the same assessment but with the nonexpert list.  Each
nonexpert was completely blind to which subset they were assessing and
the subsets were shown in a different order to each person.  The
assessed subsets were generated by GOMP-NN, ID, KM, and five different
from RAND, resulting in eight total subsets being assessed by each
participant for each list.

To compare two subsets we simply count how many features from each
list were present in the subset -- a subset which captures more of the
features will better represent the remaining elements of the dataset. 

\subsubsection{Evaluation}
We evaluate representation of each subset by reconstruction error and
human-defined feature set coverage (count). 

\begin{figure}[tb]
 \centering 
 \includegraphics[width=.9\columnwidth]{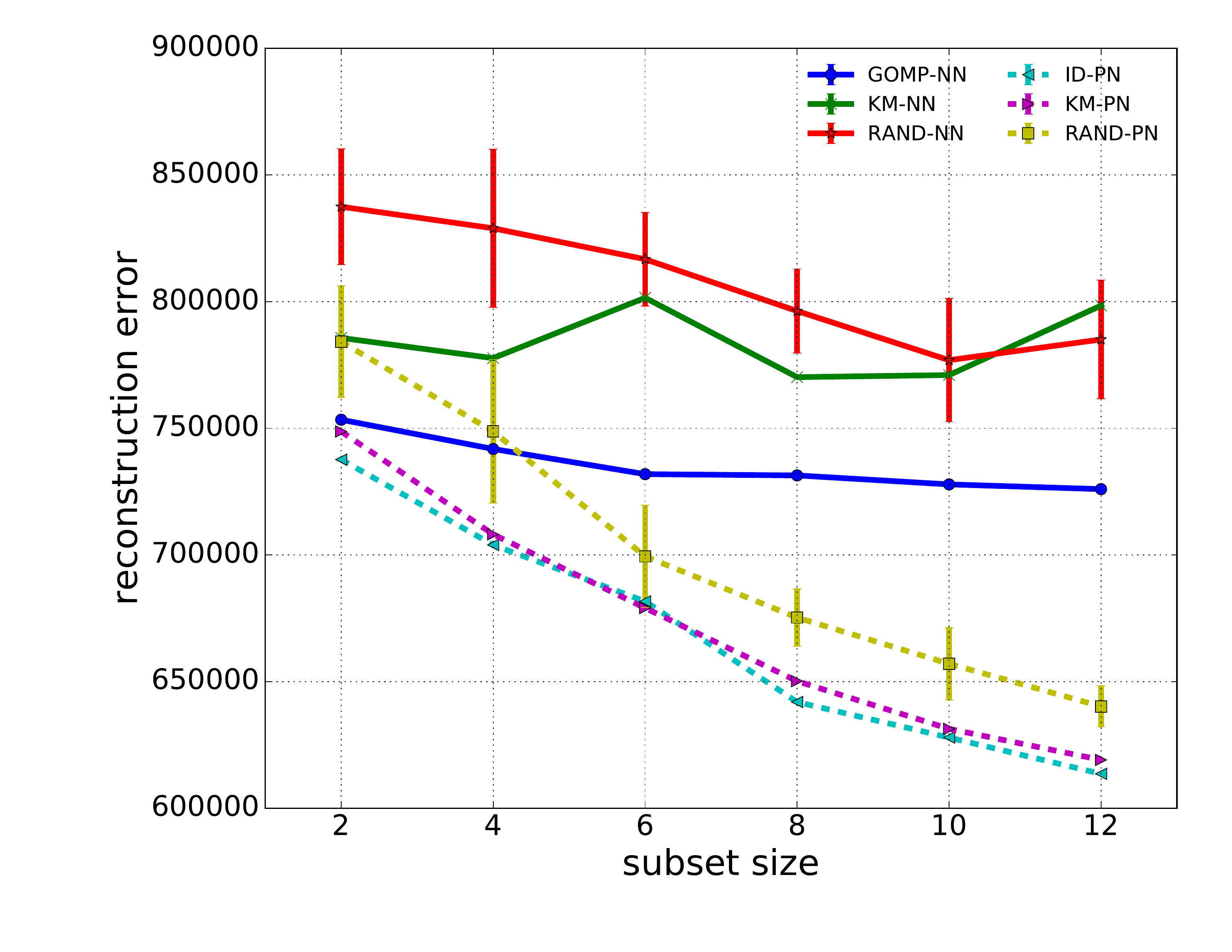} \\
 \includegraphics[width=.8\columnwidth]{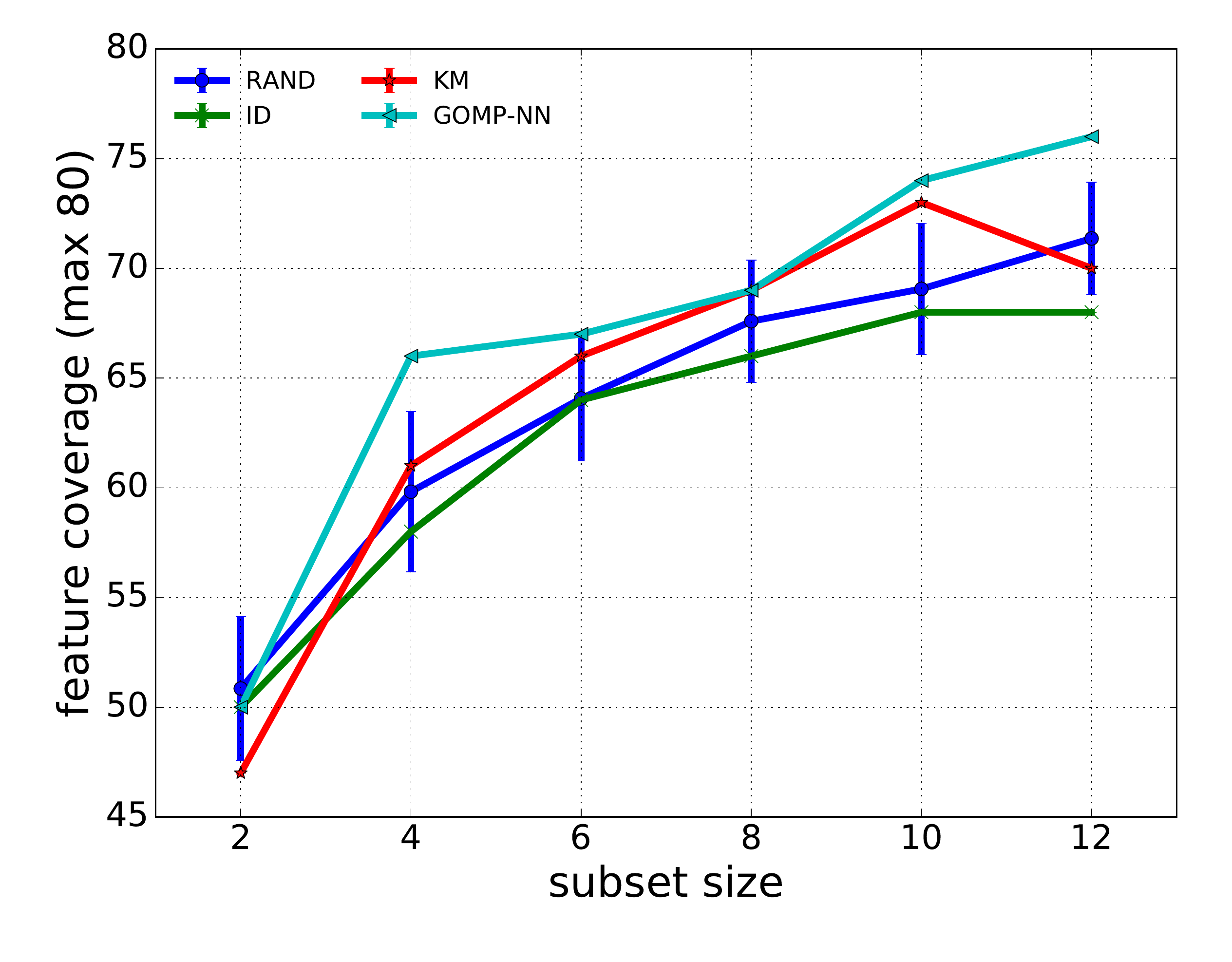}
 \caption{
\textbf{CelebA reconstruction error and coverage tests.}
	\emph{Top:} The reconstruction error for each subset type of size 2-12.
	Those using non-negative (NN) weights are shown with a solid plot line; positive-negative (PN) weighted reconstructions are shown with a dashed plot line.
    \emph{Bottom:}
	The feature count found (out of 80) in each subset type.
	The random mean and standard deviation shown were computed over $10$ and $10^2$ samples respectively per measure.
}
\label{fig:celeba:error}
\end{figure}

\Paragraph{CelebA faces dataset.}
In terms of reconstruction error using non-negative weights for the subset sizes evaluated, GOMP-NN was always shown to be more representative than RANDOM-NN and KM-NN, while KM-NN did better than random for some subset sizes.
For positive-negative weights, ID-PN and KM-PN did better than random for many sizes, but not all.
Better here means the scores was better than the mean plus one standard deviation of the corresponding RANDOM.
These quantitative representation results are shown in top of Figure~\ref{fig:celeba:error}.
However, the our comparison should also have some basis in how humans interpret the data.
To establish this connection, we also consider the human-defined feature comparison.
The results showed that GOMP-NN did better than random for all but two subset sizes, while among the others only KM really did better than random once, as shown in Figure~\ref{fig:celeba:error}.
Note that different algorithms chose very different faces to get a similar number of features.

\Paragraph{Topology optimization designs.}
The above human-face analysis showed that the proposed approach finds
a representative subset in both the reconstruction error and the more
qualitative human-defined-feature sense. 
For the specific case of STO designs, we performed a similar comparison. 
However, as described in Section~\ref{sec:userstudy}, we created the
feature list and feature assessment through a user-based evaluation
strategy. 
This limited our evaluation to a single subset size, which we chose to be $8$ because that was the maximum number of subsamples we could show on the screen without the subsample resolution becoming too low in our visual layout.
We will now summarize three types of results, a \emph{projection error} comparison, an \emph{expert-feature-list}-based assessment, and a \emph{nonexpert-feature-list}-based assessment.

In the \emph{projection-error-based} assessment we found that
GOMP-NN was the most representative of the NN methods, while KM-PN was
the most representative of the PN methods. 
In both cases all methods compared were more representative than the
average random subset error.
These results are summarized in Table~\ref{tab:shape:error}.

\begin{table}[tb]
  \caption{
	\textbf{Shape reconstruction error test results for STO D1}.
	The reconstruction error for each subset type of size 8. 
	Those using non-negative (NN) weights are shown in the top part, positive-negative (PN) weighted reconstructions are shown in the lower part.
	The random mean and standard deviation were computed over $10^2$ samples.
	The lowest reconstruction error for each group is in bold font.
	} 
	\label{tab:shape:error}
  \scriptsize%
	\centering%
  \begin{tabular}{l|ll}
  \toprule
  Method & Reconstruction error & Better? \\
  \midrule
  GOMP-NN & \textbf{628.51} & Y \\
  KM-NN & 850.44 & Y \\
  RAND-NN & 893.55(35.39) & N/A \\
  \midrule
  ID & 585.46 & Y \\
  KM-PN & \textbf{537.59} & Y \\
  RAND-PN & 609.33(26.52) & N/A\\
  \bottomrule
  \end{tabular}%
\end{table}

We also performed a similar comparison for a qualitative comparison of
human-derived features.  Rather than tag these features in all the
designs, we had the participants assess the number of features present
in each subset directly.  This was a more reasonable amount of work
for the number of participants we had, but it meant we were limited in
the number of random subsets we could evaluate.  

Even with the limited samples, we were able to show that for the \emph{expert feature list}
four of the five participants found that at least one of the nonrandom
methods did better than random.  Three of the five showed that both
GOMP-NN and ID were better than RAND, while only one of the five
showed KM-NN was better than RAND.  Overall we interpret these
results, summarized in Table~\ref{tab:shape:expert}, as support for
the methods specifically designed to select a representative set.
\begin{table}[tb]
  \caption{
	\textbf{Shape feature coverage test results using expert list for STO D1}.
	The feature count found (out of 45) from the \emph{expert feature list} for each subset type of size 8, and whether this is better or not than random.
	Here, better means more than mean plus one standard deviation.
	The random mean and standard deviation were computed over five samples given to the same participant.
	The highest feature count for each participant is in bold font.
	} 
	\label{tab:shape:expert}
  \scriptsize%
	\centering%
  \begin{tabular}{l|lllll}
  \toprule
	User & 1 & 2 & 3 & 4 & 5 \\
  \midrule
	GOMP-NN & 30          & 23          & \textbf{27} & \textbf{17} & \textbf{25} \\
	KM-NN   & 30          & 23          & 23          & \textbf{17} & 23 \\
	ID      & \textbf{31} & \textbf{26} & 26          & \textbf{17} & 24 \\
	RAND    & 27.0(4.1)   & 23.6(2.3)   & 24.6(1.9)   & 14.4(2.5)   & 21.8(1.6) \\
	\midrule
	GOMP ? & N & N & Y & Y & Y \\
	KM ? & N & N & N & Y & N \\
	ID ? & N & Y & N & Y & Y \\
	\midrule
	Any ? & N & Y & Y & Y & Y \\
  \bottomrule
  \end{tabular}
\end{table}
For the \emph{nonexpert feature list} we showed three out of four participants found an
optimal method performed better than RAND. 
Specifically three out of four for GOMP-NN, two out of four for ID, and one out of four for KM-NN.  
While these results make little distinction among the type of subset methods,
overall we take away that a method specifically designed to find a
\emph{representative} subset does in fact do better than RAND at feature
representation.  These results are summarized in
Table~\ref{tab:shape:nonexpert}.
\begin{table}[tb]
  \caption{
	\textbf{Shape feature coverage test results using nonexpert list for STO D1}.
	The feature count found (out of 17) from the \emph{nonexpert feature list} for each subset type of size 8, and whether this is better or not than random.
	The random mean and standard deviation were computed over $5$ samples given to the same participant.
	The highest feature count for each participant is in bold font.
	} 
	\label{tab:shape:nonexpert}
  \scriptsize%
	\centering%
  \begin{tabular}{l|llll}
  \toprule
	User & 1 & 2 & 3 & 4 \\
  \midrule
	GOMP-NN & 11          & \textbf{14} & \textbf{13} & \textbf{15} \\
	KM-NN   & \textbf{12} & 12          & 7           & 13 \\
	ID      & \textbf{12} & 13          & 11          & \textbf{15} \\
	RAND    & 11.2(0.8)   & 12.4(1.5)   & 7.6(0.9)    & 11.0(1.0) \\
	\midrule
	GOMP ?  & N & Y & Y & Y \\
	KM ?    & N & N & N & Y \\
	ID ?    & N & N & Y & Y \\
	\midrule
	Any ?   & N & Y & Y & Y \\
  \bottomrule
  \end{tabular}
\end{table}

\subsection{Full system evaluation}
Here we present an analysis of the full system from an STO domain
expert point of view.  For this analysis we had two domain experts
(both coauthors)
sit down and critically assess whether the tool
successfully aids in building intuition around parameters, shapes, and
quantitative design scores.  While the domain experts are coauthors,
we purposely had both play no role in the visualization
development itself, limited mainly to helping the rest of the other
authors understand what aspects of the data are important for them to
explore and suggesting some visual interactions that could be helpful.

\subsubsection{Methods}
\begin{figure}[tb]
 \centering 
 \includegraphics[width=.48\columnwidth]{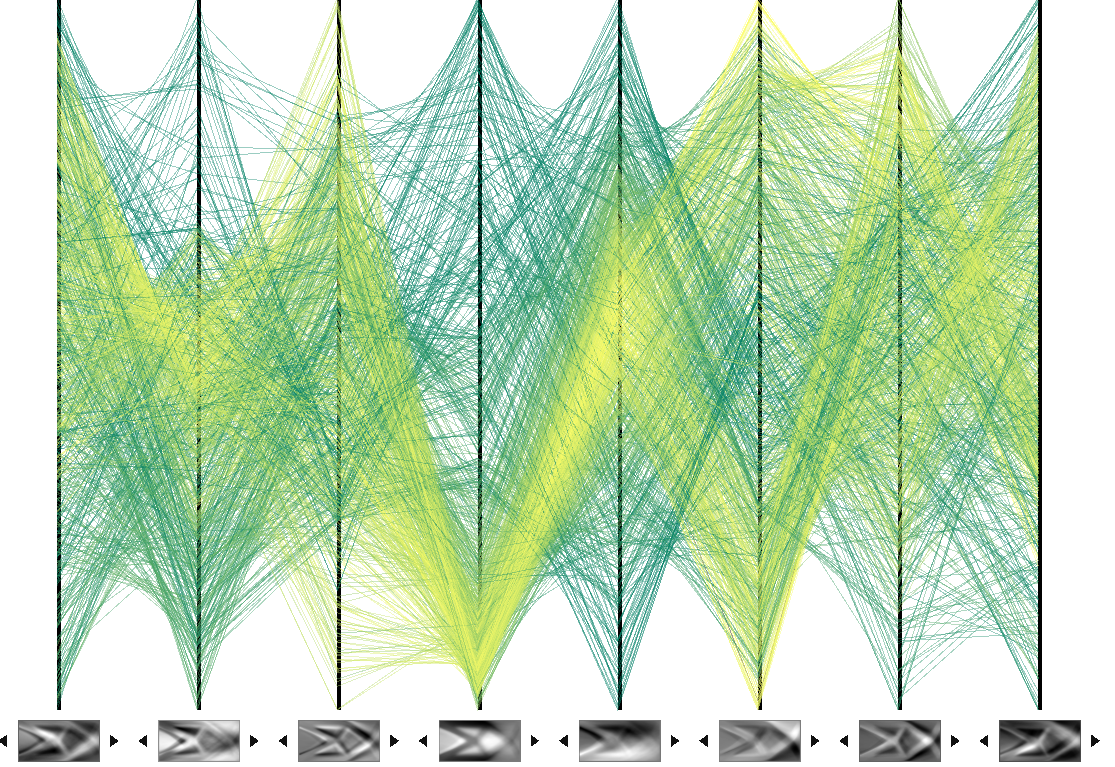} 
 \includegraphics[width=.48\columnwidth]{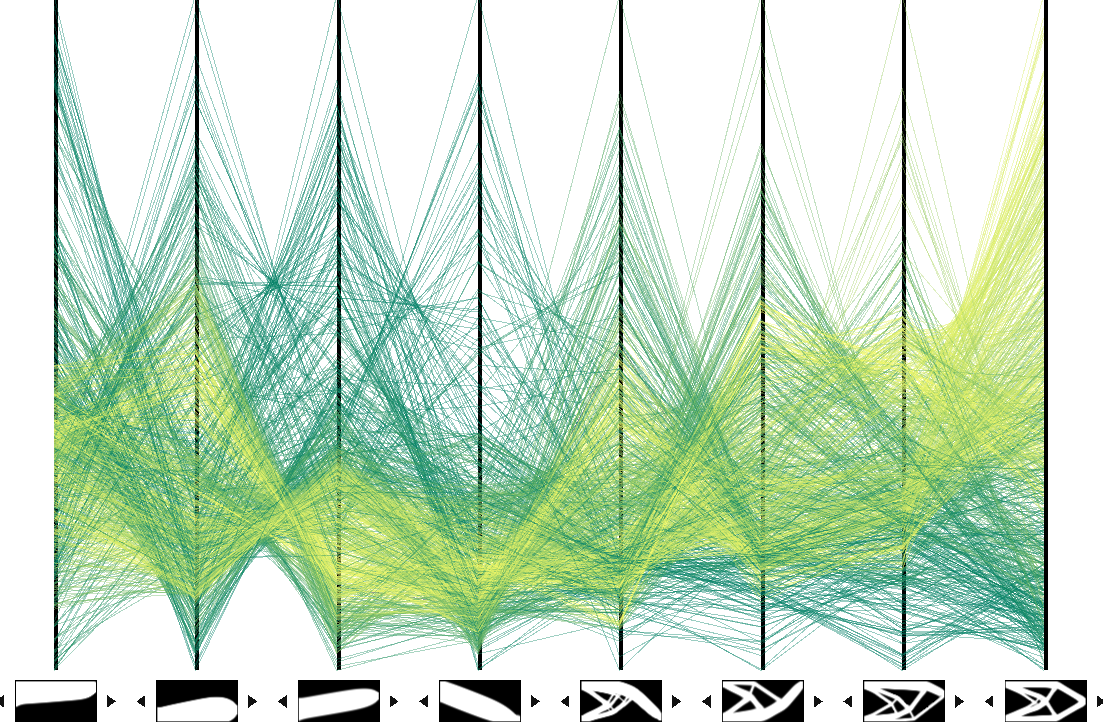} \\
	\phantom{A} \hspace{10mm} \textbf{PCA as basis}  \hfill
	\textbf{ID as basis} \hspace{10mm} \phantom{A} \\
 \caption{
\textbf{Alternatives non-negative weighted subset views.}
\emph{Left:} The same visualization approach but using a PCA basis for the axis.
Users found this approach very confusing both for the axes representation and attempting to combine them mentally using positive and negative weights.
\emph{Right:} Subset-based visualization using ID with its positive-negative weights. Users also found this combination confusing because it was difficult to manually subtract one image from another.
}
\label{fig:alternatives}
\end{figure}

We use the same set of subset-selection and weight-computation methods
explored in Section~\ref{sec:subset:eval}.  However, because manual
visual assessment is more limiting in time than the quantitative
assessments done earlier, we limited the comparison among GOMP-NN,
KM-NN, ID.  Because prior work uses PCA to analyze these type of
datasets, we also included a PCA version of the parallel coordinates
view with the appropriate positive-negative weighting, referred to as
PCA, shown in Figure~\ref{fig:alternatives}.  This provided four
different axes and weighting choices, two with non-negative weighting
and two with positive-negative weighting.

\subsubsection{Datasets}
The same parameter varying STO design dataset, \textbf{STO D1}
(described in Section~\ref{sec:subset:datasets}), was used for this
qualitative study.  We also
included a second STO design dataset, \textbf{STO D2}, where the three
parameters were fixed to $p=0$, $\theta=\pi/4$, and $\rho=1.1$
respectively, but the initial configuration was initialized randomly,
resulting in $1000$ different designs.  Because of the nonconvex
nature of the STO problem, the resulting ensemble had a large 
variety of latticed designs, although total design variance was not as great as STO D1.  The
domain experts also had much less initial intuition about this
dataset.  A sample from both datasets is shown in
Figure~\ref{fig:subsets}.


\subsubsection{Evaluation}
Before viewing the final visualization tool, we asked the two domain
experts to report any relationships they were already aware of between
parameters, designs, and scores.  We then translated those into a set
of tasks: {\em confirm each relationship known beforehand}.  The number of these prior
known relationships that can be confirmed by each version of the tool
is used as a quantitative measure for comparison.  During the case
studies, the experts were unaware of which type of subset-weight
combination was shown (labeled A,B,C), except for PCA because of the
obvious nature of the axes figures.
In addition to evaluating how many prior relationships were confirmed,
we also asked the experts to explore the data with the tool and report
any additional relationships or other insights they gained during the
use that we report as well. 

\Paragraph{Task completion assessment.}
Most of the relationships the domain experts were aware of before the
visualization focus session involved simple parameter-to-shape
relationships, such as position and contact point, angle and structure
composition (lattice vs beam), and filter size and beam thickness.
There was one relationship between shape and scoring: having many holes should lead
to high stress values.  Finally, one relationship among scores,
specifically that stress and compliance should correlate well.  One
detail to note is that there was almost no prior intuition about the
second dataset with fixed parameters and random initializations.  We
summarize the relationships identified before using the visualization
technique along with specific numbers for easy reference in
Table~\ref{tab:expert:relationships} (top portion).

While exploring the datasets the experts arrived at several unknown,
or at least unstated, relationships between the parameters and scores,
such as position and compliance, angle and compliance.  They also
confirmed some unstated assumptions, such as all designs are
variations among purely beam and purely lattice designs.  The experts
were also able to gain some intuition about the various shapes that
arose in the second dataset.  Specific relationships found in the
study are shown in the lower portion of
Table~\ref{tab:expert:relationships} with a number and asterisk to
distinguish them from the first group.

After allowing for exploration, we tasked the experts to confirm each
of the relationships they identified before using the tool.  Using the
complete tool, the experts were able to confirm all prior
relationships quite quickly with any of the basis types.

Because of the multiple components in the visualization we next tried
to understand at a finer granularity how each piece aided in the
confirmation process by removing some components.  One particular
modification that revealed a more informative comparison was to remove
the detail plot entirely, so that the expert had to rely entirely on
the parallel coordinate plot to both summarize the dataset designs and
to tease out specific details of the designs and how they relate to
the parameters and scores.  This change resulted in different performance
among the methods.

\begin{figure}[tb]
 \centering 
 \includegraphics[width=\columnwidth]{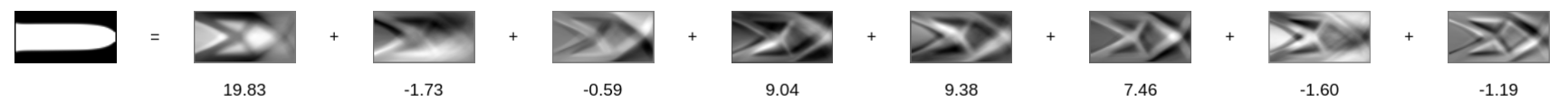} \\
	\textbf{Example of beam reconstruction using PCA} \\
 \vspace{2mm} \includegraphics[width=\columnwidth]{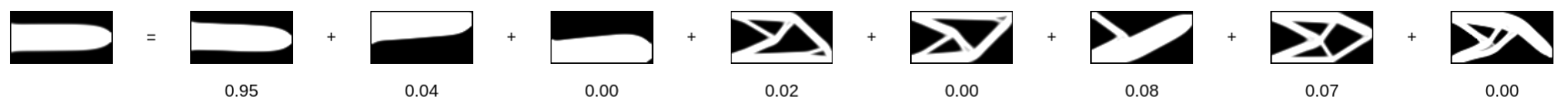} \\
	\textbf{Example of beam reconstruction using a subset} \\
 \caption{
\textbf{Beam reconstruction using PCA compared to a subset reconstruction.}
\emph{Top:} Shows a straight beam design and how it is represented using PCA modes.  
Note that none of the basis elements looks like a beam, but by combining them in positive and negative fashion the original beam can be reconstructed quite accurately.
\emph{Bottom:} The same straight beam represented by a subset.
Because some of the basis elements (subset elements) do look like a beam it is quite easy to mentally reconstruct what the data element would look like, if you did not have the design itself to view (perhaps due to space).
Note that a similar problem would occur with subset visualization if the subset did not contain a beam structure, which is another motivation for using the proposed subset selection technique by definition it is the optimal subset for reconstruction using non-negative weights.
}
 \label{fig:beam:pca}
\end{figure}

Specifically, the experts found it very difficult to confirm all but
one of the known relationships using the PCA basis or the ID basis.
When asked why it was difficult, they were quite critical of the PCA
basis set and weights, as it was ``impossible to combine the bases,
especially with the positive-negative weights''.  In essence, the
experts found it very difficult to mentally combine the PCA basis to
understand any of the reconstructions.  One particularly illuminating
example is that none of the PCA bases showed an example of a beam,
because all beam representations were recreated by subtracting
lattice-looking bases from each other; an example is shown in
Figure~\ref{fig:beam:pca}.  The ID subset visualization was also
difficult for the experts to use in confirming relationships.  Even
though the axes were a subset of the designs, the positive-negative
weights made the mental combination of examples difficult.   This
confirms our hypothesis, which is that {\em constructive combinations of
features allows for easier interpretation of data represented as
mixtures of basis elements.  }

In contrast, the experts were able to confirm most of the prior
relationships using KM-NN and GOMP bases.  Consider the example
comparison in Figure~\ref{fig:beam:pca} where mentally extracting a
beam design from the PCA basis is quite difficult, in contrast to the
very obvious beam structure indicated by the subset visualization
example.  For some of the relationships, experts claimed a partial
confirmation because of a desire to view more samples of some of the
more detailed aspects of the dataset than those available.  That experience
outlines one limitation of the approach -- that while the
subset-based view provides a good overview of the dataset with feature
details, it is difficult to extrapolate out to feature details not
represented in the subset.

Most of this discussion has centered around the first of the two STO datasets, but
results were also investigated for the second dataset.  
Most of the prior assumptions were not applicable to the second
dataset because the parameters were fixed, but for the two that did
apply we observed similar results.  The precise results are summarized
in Table~\ref{tab:expert:verify}.

\begin{table}[tb]
  \caption{\textbf{Relationships reported by domain experts}.
	The relationships reported before viewing the visualization tool are summarized in the top portion of table with assigned numbers.
	The relationships reported (discovered) while using the tool are reported in the lower half of the table (numbers with asterisks).
	} 
	\label{tab:expert:relationships}
  \scriptsize%
	\centering%
  \begin{tabular}{l|l}
  \toprule
	relationship & description \\
  \midrule
	1 & position and right side contact point\\
	2 & angle horizontal then structure horizontal\\
	3 & low angle and varying positions then shallow \\
	  & angle straight structure\\
	4 & filter size controls structure thickness\\
	5 & more holes in structure then higher stress\\
	6 & compliance and stress are correlated\\
	\midrule
	7* & symmetric-quadratic relationship between \\
	 & position and compliance/stress\\
	8* & position choice induces an upper limit \\
	 & on compliance (lowest at 0) \\
	9* & negative correlation between angle and compliance \\
	10* & all of the designs are variations of a beam or \\
	 & lattice design (confirmation)\\
  \bottomrule
  \end{tabular}
\end{table}

\begin{table}[tb]
  \caption{\textbf{Prior known relationships assessments by domain experts}.
	Verification of known relationships described in Table~\ref{tab:expert:relationships} using various versions of the visualization tool (by changing axis and visualization components).
	\emph{Detail} indicates use of visualization with detail view, otherwise the detail view was excluded.
	Results are shown for the first (STO D1) and second (STO D2) datasets in upper and lower parts of the table, respectively.
Here Yes (Y) means confirmed, No (N) cannot confirm, Partial (P) means confirmed but want more samples to be completely sure.
	} 
	\label{tab:expert:verify}
  \scriptsize%
	\centering%
  \begin{tabular}{l|llllll}
  \toprule
	relationship&1 &2 &3 &4 &5 &6\\
  \midrule
	Detail (D1) & Y& Y& Y& Y& Y& Y\\
	GOMP-NN (D1)& Y& P& Y& P& P& Y\\
	ID (D1)     & N& N& N& N& N& Y\\
	KM-NN (D1)  & P& P& P& N& P& Y\\
	PCA (D1)    & N& N& N& N& N& Y\\
	\midrule
	Detail (D2) & -& -& -& -& Y& Y\\
	GOMP-NN (D2)& -& -& -& -& P& Y\\
	ID (D2)     & -& -& -& -& N& Y\\
	KM-NN (D2)  & -& -& -& -& P& Y\\
	PCA (D2)    & -& -& -& -& N& Y\\
  \bottomrule
  \end{tabular}
\end{table}

\subsubsection{Case studies} \label{sec:experts} 
We now present several case studies from the domain expert session.
These case studies detail some of the uses we found for
the tool, and identify some of the specific strengths and limitations
of the proposed subset-based visualization.\vspace{0.01in}\\ 
\noindent\textbf{Subset-based parallel coordinate view as surrogate to full dataset.}
As part of the relationship confirmation task above, we forced the
domain experts to not use the detail view (individual elements), and
the resulting use of the visualization was quite illuminating.  When
limited to the the subset-based parallel coordinate view, the experts
selected points in the parameter scatter plot and then looked at which
of the subset elements contained large weights.  For the case of
non-negative weighted views, they used those values to mentally
consider what the element looks like and make a conclusion.  This
subset-based view usage is in contrast to the individual element view,
where the experts selected points from the parameter scatter plot and
looked at the corresponding element.  To summarize, the subset-based
view was being used as a surrogate for the full dataset in completing
the confirmation tasks mentioned above.  Note that in this same
setting when the subset-based view was changed to use
positive-negative weights or the PCA basis elements, this summarizing
ability of the parallel coordinate view was apparently lost.
In considering this aspect of the subset-based parallel coordinate view, the domain experts commented that it ``Allows for a whole-view perspective of the data, instead of viewing the data elements individually''.\vspace{0.01in}\\
\noindent\textbf{Limitations of subset-based view.}
One of the tasks was to confirm that the filter size influenced the
size of the structures in the design (smaller filter size results in
smaller lattice structures and other features).  However because the
subset contains only a limited number of the smaller lattice structure
examples, the expert was unable to confirm the relationship fully,
marking it as ``partially'' confirmed.  Because the experts had
previously confirmed the relationship using individual instances, they
were expecting to see more examples of the relationship than what was
available.  The experience highlights one limitation of the method:
when any feature is not included in the subset, the surrogate nature
of the view is not nearly as true to the underlying dataset.
This point emphasizes that the choice of subset becomes crucial to
success, and why we advocate the use of the GOMP-NN-based
selection.\vspace{0.01in}\\ 
\noindent\textbf{Confirmation of design shape aspect of
  parameter-score relationships.}
The domain experts were able to discover new relationships among
parameters and scores, for example relationships 7*,8*, and 9*, in
Table~\ref{tab:expert:relationships}.  These relationships have
implications for the design shapes themselves, and in each case the
experts were able to use the subset-based view to understand that
relationship.
For example, for relationship 9* they were able to show that the
steeper angle resulted in a more latticed structure that had more
holes and higher stress values.\vspace{0.01in}\\ 
\noindent\textbf{Exemplar scatter plot as summary device.}
The exemplar scatter plot, although not used extensively by the
experts, did become useful for understanding and confirming general
trends in the dataset, as well as for understanding the parallel
coordinates plot at times.  For example, the domain experts were able
to confirm that the STO designs can be largely grouped into two large
sets, those that have a large central beam, and those that are mainly
lattice structure.
This confirmation was accomplished by viewing the \emph{exemplar
  scatter plot} with a beam structure on one axis, and a purely
lattice structure on the other axis.\vspace{0.01in}\\ 
\noindent\textbf{Parameter scatter plot as driver for discovery.}
For the full tool, the parameter-based scatter plots (i.e., scatter
plot of angle and compliance) that is then colored by a third variable
such as maximum stress, proved to be the primary tool used to drive
discovery.
The basic workflow consists of the experts looking at relationships on
that plot, and then using the linked-nature of the views to explore
visual aspects of the relationships in the subset-based visualization
and the detail view.\vspace{0.01in}\\ 
\noindent\textbf{Limitations of PCA- and ID-based views.}
The limitations of the PCA view became very evident in the domain expert study.
While using the subset view, they were able to confirm and gauge
general visual trends in the data, but when using the PCA version of
the tool the users became quite frustrated. 
Some of the phrases used to describe their reaction are: ``I can't see any beams in the modes'', ``convoluted examples... look like photo negatives... or ghosts of the structures'', and ``the axis are no longer intuitive''.
The positive-negative weights also played a role in the confusion.
This was discovered by using the ID-based subset view, that uses
subsets but with positive-negative weights. 
While the axes were easier to understand for the users, trying to
mentally subtract examples from each other was difficult to do. 
When the bases were PCA modes, the task became even more difficult.

\subsubsection{Additional examples} \label{sec:clarify}
To further discuss some of the points made above, and to further demonstrate how the system benefits from the use of a subset-based display of the designs, we will now present some examples of the more challenging \emph{STO2} datasets.
These example were not included in the case study, but are presented here for further illumination of the points brought up previously.
Note that the STO2 dataset mentioned above, used as part of the expert case study, is more challenging to understand, even for an expert, because the only parameter that was changed was the initial material placement.
In other words, the angle, position, and filter size parameters are all constant.
This makes the dataset difficult to interpret because there is no longer a clear connection between parameter choice and the resulting STO design.
However, we will show that the system is still quite useful for understanding the connection between attributes, such as stress, and the visual design.

\begin{figure*}[tb]
 \centering 
 \includegraphics[width=0.45\textwidth]{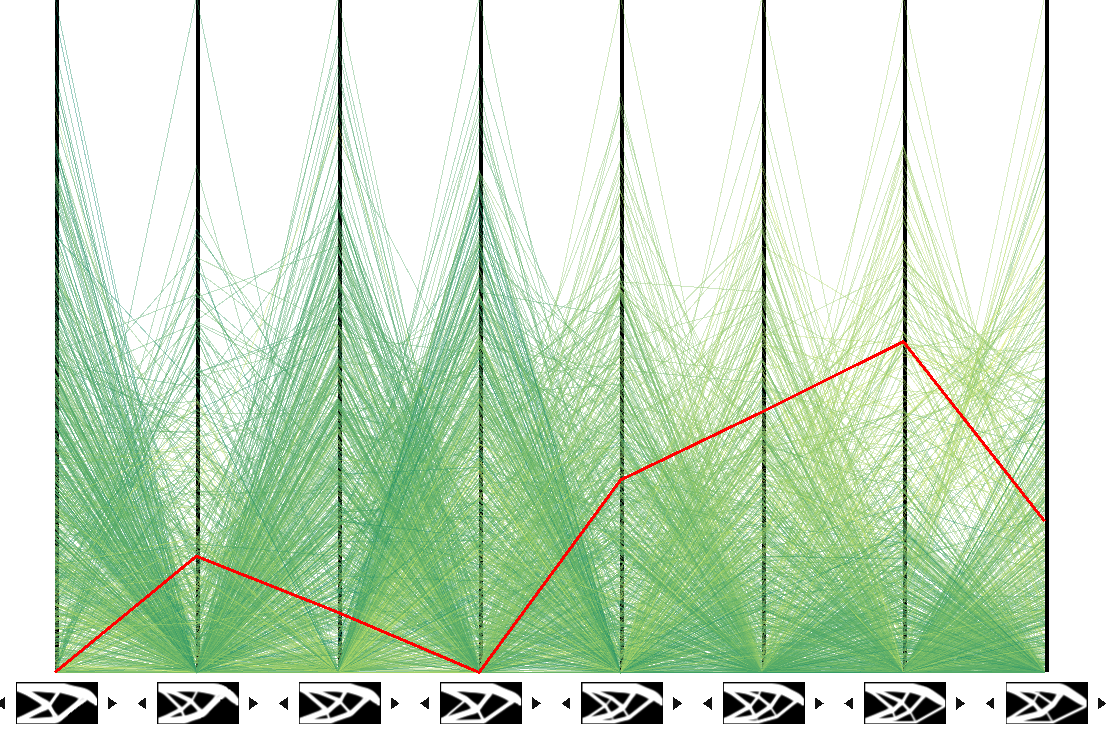} 
 \includegraphics[width=0.45\textwidth]{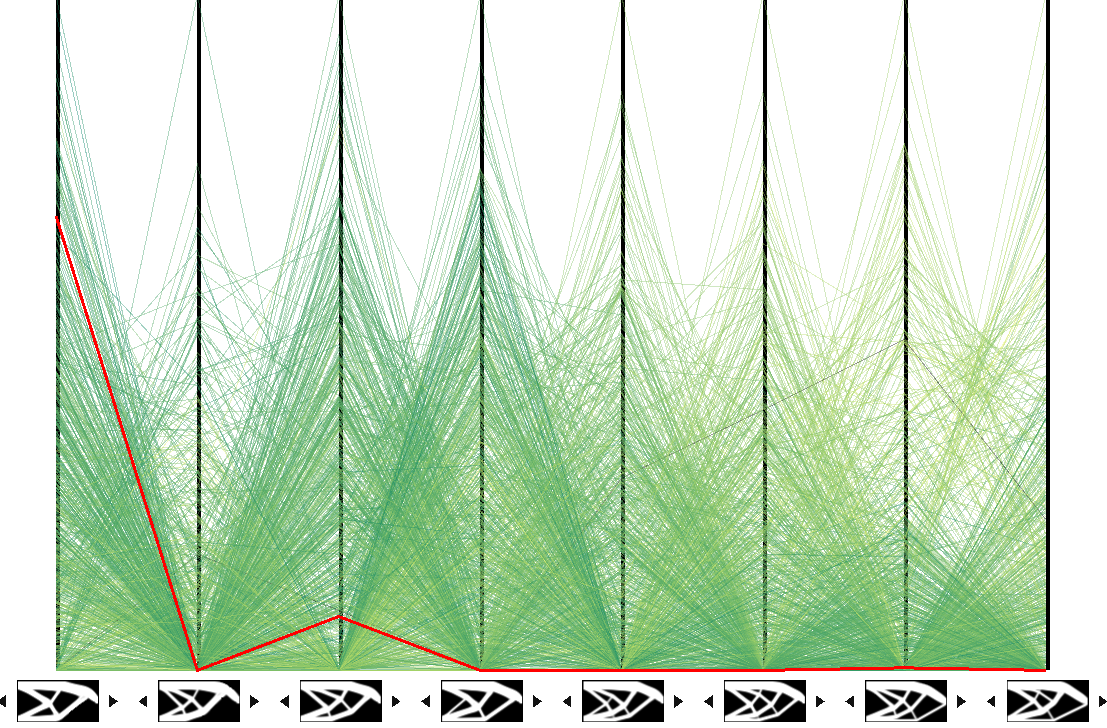} \\
	\phantom{A} \hspace{30mm} \textbf{High stress example selected} \hfill
	\textbf{Low stress example selected} \hspace{30mm} \phantom{A} \\
 \caption{
\textbf{GOMP-NN view of the more challenging STO2 dataset.} This example shows how visual features can be directly connected to low and high stress using a subsample-based view (more lattice structure correlates with high stress).
}
 \label{fig:sto2:gomp}
\end{figure*}

The subset-based summary is shown in Figure~\ref{fig:sto2:gomp} where the STO2 dataset has been loaded. 
The display is using the GOMP-NN basis and weights with a high stress example selected and highlighted in the left figure, and a low stress figure has been highlighted in the right figure.
Because the subset-based view captures and shows the a subset of the dataset, which in this case appears to be the differing lattice structure, it becomes quite clear from the visualization that the more lattice structure there is in the design the higher the stress becomes.

\begin{figure*}[tb]
 \centering 
 \includegraphics[width=0.45\textwidth]{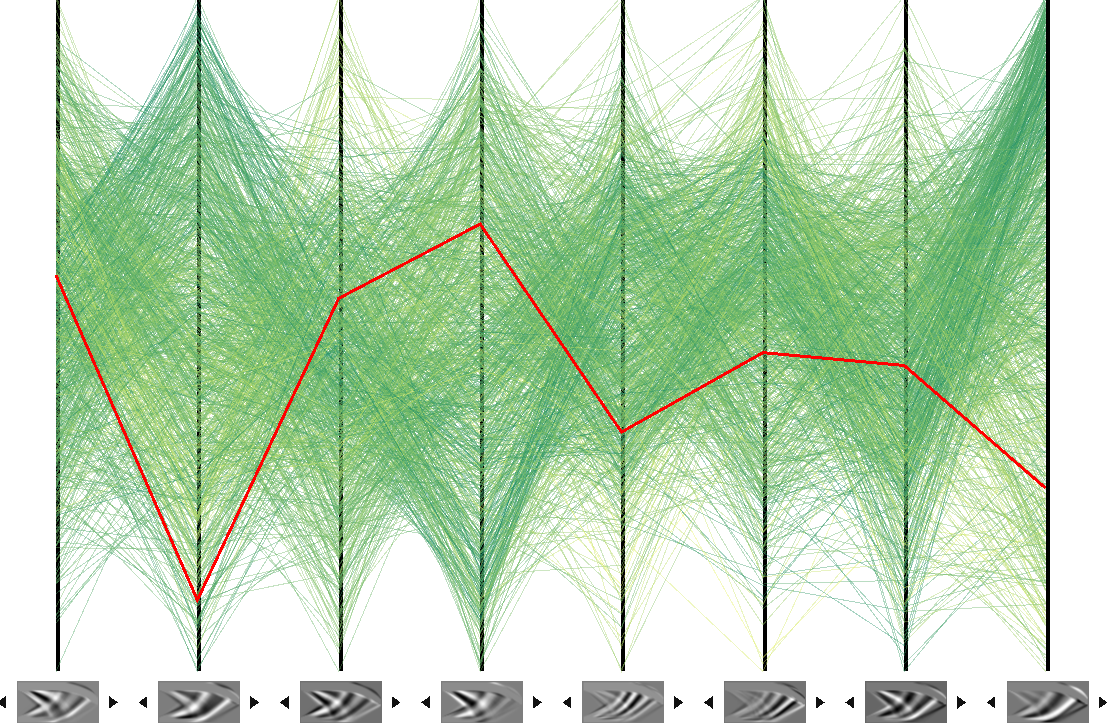} 
 \includegraphics[width=0.45\textwidth]{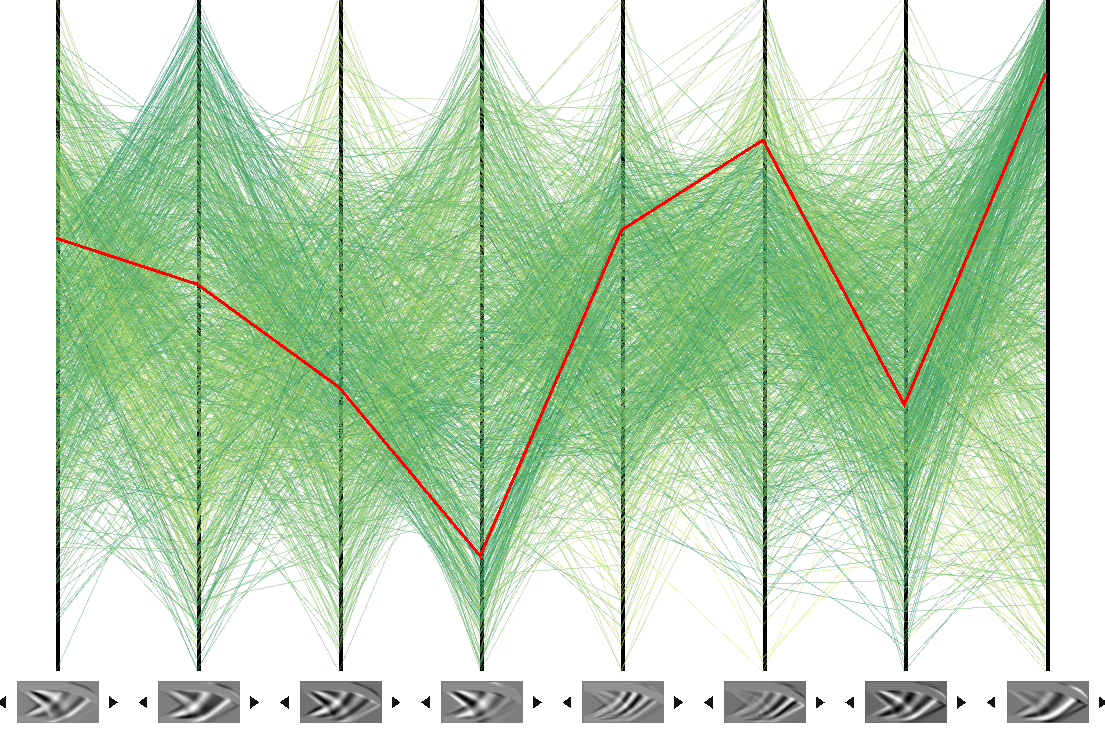} \\
	\phantom{A} \hspace{30mm} \textbf{High stress example selected} \hfill 
	\textbf{Low stress example selected} \hspace{30mm} \phantom{A} \\
 \caption{
\textbf{PCA view of the more challenging STO2 dataset.} This example shows how it is difficult to find a connection between low and high stress and the design features using a PCA-based view.
}
 \label{fig:sto2:pca}
\end{figure*}

Now contrast the very clear interpretation of that observation with what we see in the PCA-based view shown in Figure~\ref{fig:sto2:pca}.
Again we show the same high stress example in the left part of Figure~\ref{fig:sto2:pca} and the same low stress example in the right part.
Now, because of the choice of PCA basis, it is quite difficult to understand what, if any, relationship can be seen using this view of the data.

This example reinforces what we observed on other datasets, such as STO1, CelebA, and the sketch face database.
In terms of interpretability, a subset-based view provides superior summarization to PCA and related basis choices.


\section{Discussion}

\Paragraph{Strengths.}
The representative subset view appears to be an appropriate technique for summarizing a dataset of visual features, and specifically for STO designs.
By linking the subset-based view to the STO parameters through coloring and interactive selection, the visualization allows for a parameter-driven exploration of the dataset while the subset view provides the needed dataset visualization, avoiding the need to display all individual designs for most tasks.
The method also appears suitable for more general datasets with visual features, and potentially non-visual features given appropriate linked views.

\Paragraph{Limitations.}
Extrapolation of any features not present in subset is difficult to make, which can limit the amount of detailed relationships that can be found using the subset view.
Additionally, our proposed method is limited to design spaces where linear combinations work. 
While we have found this to work for STO designs, some design spaces will require a different approach.
In terms of our assessment, we limited ourselves (on purpose, for matters of scope) to 2D design spaces, but in the future it would be interesting to consider higher dimensional design spaces.
While we explored several aspects of the method, our assessment was primarily based around the STO design dataset and the CelebA image database. 
In the future we would like to explore how this approach benefits in other data types, such as text and other nonspatial datasets.

\Paragraph{Conclusions.}
The proposed visualization method has significant advantages over traditional dimension-reduction-based techniques in terms of allowing a domain expert to extrapolate and infer details of the dataset from a small, carefully chosen subset.
For STO design this specifically allows connecting parameters and score values back to visual design features in many cases.
The suggested optimally representative GOMP-NN-based subset selection has been shown to be the most representative in terms of reconstruction error for non-negative weighting and, in a more qualitative sense, more representative of human-determined features than a random dataset.
Finally, expert case-studies have identified some of the strengths and limitations of the proposed approach.

\ifCLASSOPTIONcompsoc
  \section*{Acknowledgments}
\else
  \section*{Acknowledgment}
\fi

D.~Perry and V. Keshavarzzadeh acknowledge that their part of this research was sponsored by ARL under Cooperative Agreement Number W911NF-12-2-0023. S.~Elhabian, R.M.~Kirby and R.~Whitaker acknowledge support from DARPA TRADES HR0011-17-2-0016.
The views and conclusions contained in this document are those of the authors and should not be interpreted as representing the official policies, either expressed or implied, of ARL, DARPA or the U.S. Government. The U.S. Government is authorized to reproduce and distribute reprints for Government purposes notwithstanding any copyright notation herein. 

\ifCLASSOPTIONcaptionsoff
  \newpage
\fi



\bibliographystyle{IEEEtran}
\bibliography{main}



\end{document}